\documentclass[11pt,letterpaper,nofootinbib,eqsecnum,floats,floatfix,amsmath,superscriptaddress,amssymb]{revtex4-1}
\pdfoutput=1
\usepackage{verbatim}
\usepackage[dvipsnames]{xcolor}
\usepackage{bm}
\usepackage{latexsym}
\usepackage{mathrsfs,blkarray}
\usepackage{multirow}
\definecolor{darkgray}{RGB}{25,25,25}
\usepackage[%
             colorlinks=true,urlcolor=MidnightBlue,linkcolor=MidnightBlue,citecolor=OliveGreen,
             pdfpagelabels=true,hypertexnames=true,
            plainpages=false,naturalnames=false,
             ]{hyperref}
\usepackage{graphicx}
\usepackage{dcolumn}
\usepackage{amsfonts}

\begin{document}

\renewcommand{\baselinestretch}{1.3}
\parskip 0.05in

\newcommand{\qui}{$\mathbf{Q}\;$} 
\newcommand{\ver}{$\mathbf{Q}_0\;$}
\newcommand{\edg}{$\mathbf{Q}_1\;$}
\newcommand{\pair}{$(\mathbf{Q}_0,\mathbf{Q}_1)\;$}
\newcommand{\Tr}{{\rm Tr}}
\newcommand{\que}{\textbf{!! QUESTION !!}}
\newcommand{\todo}[1]{{\bf ?????!!!! #1 ?????!!!!}\marginpar{$\Longleftarrow$}}
\renewcommand{\comment}[1]{}

\renewcommand{\thefootnote}{\fnsymbol{footnote}}

\newcommand{\setall}{\setcounter{equation}{0}}
\newtheorem{definition}{\sf DEFINITION}
\newtheorem{conjecture}{\sf CONJECTURE}
\providecommand*\email[1]{\href{mailto:#1}{#1}}

\def\lsim{\mbox{\raisebox{-.6ex}{~$\stackrel{<}{\sim}$~}}}
\def\gsim{\mbox{\raisebox{-.6ex}{~$\stackrel{>}{\sim}$~}}}
\def\lsim{\mbox{\raisebox{-.6ex}{~$\stackrel{<}{\sim}$~}}}
\def\gsim{\mbox{\raisebox{-.6ex}{~$\stackrel{>}{\sim}$~}}}
\renewcommand{\t}{\tilde}

\noindent \begin{normalsize}KCL-PH-TH/2012-46 \\Imperial/TP/12/AH/04 \end{normalsize}

\title{\Large \bf Superconformal Block Quivers, Duality Trees and Diophantine Equations}

\author{Amihay Hanany}
\email{a.hanany@imperial.ac.uk}
\affiliation{Theoretical Physics Group, The Blackett Laboratory,
Imperial College, Prince Consort Road, London SW7 2AZ, UK}

\author{Yang-Hui He}
\email{hey@maths.ox.ac.uk}
\affiliation{Department of Mathematics, City University, London,
Northampton Square, London EC1V 0HB, UK}
\affiliation{Merton College, University of Oxford, OX1 4JD, UK}
\affiliation{School of Physics, NanKai University, Tianjin, 300071, P.R.~China}

\author{Chuang Sun}
\email{chuang.sun@physics.ox.ac.uk}
\affiliation{School of Physics, NanKai University, Tianjin, 300071, P.R.~China}
\affiliation{Rudolf Peierls Centre for Theoretical Physics,
1 Keble Road Oxford OX1 3NP}

\author{Spyros Sypsas}
\email{spyridon.sypsas@kcl.ac.uk}
\affiliation{Department of Physics, King's College London,
Strand, London WC2R 2LS, UK}


~\\

\begin{abstract}
\vspace{1.5cm}
\begin{center}{\bf Abstract}\end{center}

We generalize previous results on $\mathcal{N}=1$, $(3+1)$-dimensional superconformal block quiver gauge theories. It is known that the necessary conditions for a theory to be superconformal, i.e. that the beta and gamma functions vanish in addition to anomaly cancellation, translate to a Diophantine equation in terms of the quiver data. We re-derive results for low block numbers revealing an new intriguing algebraic structure underlying a class of possible superconformal fixed points of such theories. After explicitly computing the five block case Diophantine equation, we use this structure to reorganize the result in a form that can be applied to arbitrary block numbers. We argue that these theories can be thought of as vectors in the root system of the corresponding quiver and superconformality conditions are shown to associate them to certain subsets of imaginary roots. These methods also allow for an interpretation of Seiberg duality as the action of the affine Weyl group on the root lattice. 
\end{abstract}

\maketitle
\hypersetup{linktocpage}
\tableofcontents


\section{Introduction}
Over the last few decades, the study of quiver theories has occupied a 
prominent position both in pure mathematics, especially in algebraic 
geometry and representation theory 
(cf.~e.g.,\cite{Crawley,Derksen,Savage,Brion,Assem:book}), and in 
theoretical physics, especially in the AdS/CFT correspondence and in the 
phenomenology of Standard-like models 
(cf.~e.g.,\cite{Douglas:1996sw,Benvenuti:2004dw,He:1999xj,Hanany:1998sd,Berenstein:2006pk}).
One salient feature is that gauge theories arising as world-volume 
quantum field theories living on stacks of branes probing Calabi-Yau 
singularities naturally have a product structure for the gauge group as 
well as bi-fundamental and adjoint fields realized by open-strings;
such generically supersymmetric gauge theories are thus encoded by quivers.

The dialogue between the world-volume physics and the geometry of the 
Calabi-Yau singularity has given us a wealth of new physics and 
mathematics over the last score of years.
There is a variety of such theories one can construct, or 
``geometrically engineer'',  in this way depending on the type of branes 
and the choice of the Calabi-Yau space.
Of main interest has been the construction of $(3+1)$-dimensional gauge 
theories preserving $\mathcal{N}=1$ or $\mathcal{N}=2$ supersymmetries, 
which feature centrally to the $AdS_5/CFT_4$ correspondence 
\cite{Maldacena:1997re} and which, of course, are of some 
phenomenological concern.
There has been an industry to 
construct even more classes of such quiver gauge theories with an 
underlying geometry, ranging from orbifolds 
\cite{Douglas:1996sw,Johnson:1996py,Hanany:1998sd}, to toric 
singularities 
\cite{Douglas:1997de,Beasley:1999uz,Feng:2000mi,Feng:2001bn}, as well as 
their avatars as brane tilings
\cite{Hanany:2005ve,Franco:2005rj,Franco:2005sm,Feng:2005gw,Hanany:2012hi}, to more 
generic spaces \cite{Cachazo:2001gh,Wijnholt:2002qz,Feng:2007ur}.
A myriad of theories have been established and countless successes, 
recounted.

Let us focus on ${\mathcal N}=1$ theories.
Indeed, whereas the ${\mathcal N}=2$ Lagrangian is fixed once the matter 
content is specified, whereby limiting the possibilities for 
interaction, the ${\mathcal N}=1$ superpotential is an additional 
ingredient to the matter specified by the quiver.
In fact, the F-terms prescribe formal algebraic relations to the arrows 
in the quiver, giving rise to so-called labelled quivers with relations, 
which has been recently intensely investigated by mathematicians.
Furthermore, a key advantage of ${\mathcal N}=1$ is chirality - a 
desired phenomenological property; in terms of the quiver, this is 
reflected by the fact that not every arrow between two nodes has a 
counter-part going in the opposite direction.
Finally, because of the inherent holographic nature of certain classes of our gauge 
theories, they have superconformal fixed points in the infra-red.
This is, of course, reflected by the archetypal example of AdS/CFT, the 
${\cal N}=4$ super-Yang-Mills theory in $(3+1)$-dimensions from which 
all our quiver theories geometrically descend.

The natural question thus arises as to whether one could march toward a 
classification scheme of the plethora of superconformal ${\mathcal N}=1$ 
quiver gauge theories which have bedecked the literature. 
This is, of course, an ambitious goal, especially given the unclassified 
nature of Calabi-Yau threefold singularities. Note though that the quivers studied here
are more general since they are not necessarily Calabi-Yau threefolds.
In the toric subclass of Calabi-Yau manifolds, due to the combinatorial 
nature of the geometry, attempts are under way towards an enumeration 
\cite{Hewlett:2009bx,Davey:2009bp,Hanany:2012hi,Hanany:2010cx,Hanany:2010ne}.

The organization of quivers by grouping nodes which are unlinked into 
so-called ``blocks'' has emerged in the study of sheaves over del Pezzo 
surfaces \cite{1997alg.geom..3027K}.
This was also applied to the superconformal context over the years 
\cite{Herzog:2003dj,Herzog:2003zc,Herzog:2004qw,Feng:2002kk,Aspinwall:2004vm}, 
culminating in a systematic investigation in \cite{Benvenuti:2004dw}.
Such seemingly innocuous procedure turns out to be very powerful.
As demonstrated in \cite{Benvenuti:2004dw}, many of the known 
theories, often corresponding to such complicated geometries as cones over 
Hirzebruch surfaces or pseudo del Pezzo surfaces, can have their quiver 
diagrams contracted to ones with only a few blocks. Moreover, the necessary 
conditions for such a theory to be superconformal, i.e. vanishing beta and gamma 
functions, translate to Diophantine equations over the quiver data \cite{Feng:2002kk,Franco:2003ja}.

Now, it had been realized that Seiberg duality is a very particular 
transformation on quiver theories 
\cite{Feng:2000mi,Feng:2001bn,Beasley:2001zp} and various geometrical 
interpretations ranging from Picard-Lefschetz transformations 
\cite{Feng:2002kk} and Weyl group action on the quiver root system \cite{Cachazo:2001sg,Fiol:2002ah}, 
to mutations in exceptional collections of coherent 
sheaves \cite{Herzog:2004qw} and to tiltings in the derived category 
\cite{Berenstein:2002fi,Braun:2002sb} have been studied.
Such a duality is well adapted to the block structure. The possible values, 
after a blossoming ``tree" of duality transformations, all satisfy the 
Diophantine equation determined by the geometry. In other words, the Diophantine equation is
an invariant of Seiberg duality.

Our motivation is clear:
First, we wish to continue the study of the taxonomy of ${\cal N}=1$ 
quiver theories, organized by blocks.
In \cite{Benvenuti:2004dw}, the situation up to four-blocks was detailed.
The reason the case study stopped there is because starting from 
five-blocks, a qualitative difference arises:
it is not clear which cycles enter in the superpotential, and it is not 
completely clear if an arbitrary number of Seiberg Dualities leave the 
quiver chiral.
Our first challenge is to address this issue in a completely algorithmic 
and exhaustive way. Indeed, in Sec. \ref{s:5b} we will see that Seiberg duality 
leaves the models chiral.

This possibility to continue to a higher number of block is only the tip 
of the iceberg.
We shall see how the representation theory of quivers comes to our aid 
and offers us a unifying light under which we could examine the quiver 
block structure, the assignment of ranks and arrows, as well as the 
general form which the Diophatine equations must assume.
Thus representation theory, algebraic geometry and number theory come 
into full interplay with the physics.

The paper is organized as follows:
We begin in Section \ref{s:block} by setting the notation of our problem 
of classifying block-quivers by illustrating with the known examples of 
three and four-blocks.
Along the way, we reveal a new structure of the block models and show how
the Diophantine equation can be written as a sum over minors of the quiver 
adjacency matrix. Using the representation theory of quivers, especially a 
certain bi-linear form called the Tits form, we reveal the origin of this 
formula and show how Seiberg duality is realized 
in this context. The only input for the derivation of the superconformal condition
is the quiver matrix, a fact implying, quite surprisingly, that the Tits form
automatically encodes the vanishing beta and gamma functions of the theory. 
We also derive a similar formula for the four-block quiver and
describe its reduction to the three block case.
Then, in Section \ref{s:5b} we study the first non-trivial case of 
five-block quivers, which have eluded much of the physics and 
mathematics literature.
We show, despite the complicated combinatorics, that we can still 
use the Tits form to organise the Diophantine equation and shed light 
into Seiberg duality.
We conclude with outlooks in Section \ref{s:conc}.
Of use will be Appendix \ref{ap:quiver} which is an enlightening but 
self-contained review of the rudiments of quiver representation theory 
which will be used in the paper.

\section{Block Quivers}\label{s:block}\setall
In this section we begin with a brief reminder for the reader of the concept of block quivers, how Diophantine equations arise from the requirement of existence of superconformal fixed points, as well as the emergence of representation-theoretic quantities in relation to physical constraints.
We will illustrate with the well-known example of the three-block quivers, under a new and unifying light.
For short, self-contained exposition to some relevant terminology of quivers, especially from a mathematical perspective, we refer the reader to Appendix \ref{ap:quiver}.

\begin{figure}[tbh]
\centering
\begin{tabular}{cc}
\includegraphics[scale=0.25]{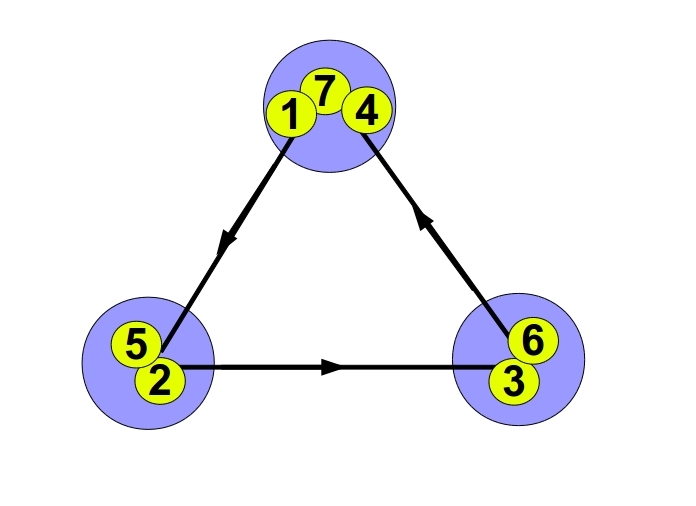} &
\includegraphics[scale=0.25]{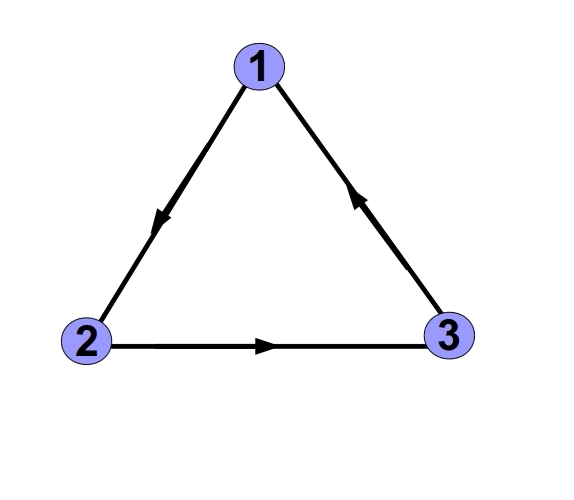} 
\end{tabular}
\caption{{\sf A three-block structured quiver diagram. Quiver nodes in yellow are gathered in block nodes in violet. The arrows in both pictures denote collectively all possible arrows among the indicated yellow nodes.}
}
\label{fig:3b}
\end{figure}

The central object of our concern is the {\it chiral quiver}, by which we mean any quiver diagram which has no bi-directional arrows (including, in particular, self-adjoining loops which connect a node to itself) and in addition all fields between two adjacent vertices have the same R-charge. 
The reason for this restriction will soon be clear; essentially it is because we will only be dealing with anti-symmetrized adjacency matrices which do not capture the information of bi-directional arrows. The constraint of no bi-directional arrows is not severe as the majority of the myriad of quivers which have risen over the last decade of the AdS/CFT correspondence belongs to this category. The requirement of equal R-charges among fields charged under adjacent blocks though, does restrict our treatment. Nevertheless, we will find that known five block del Pezzo quivers are still solutions of our Diophantine equation. In these cases though the R-charges can not be computed using our methods and one has to use standard tools like a-maximization \cite{Intriligator:2003jj}. 
Now following \cite{Benvenuti:2004dw}, we recall that a {\it block} in a chiral quiver diagram is as follows: 
\begin{definition}
A block is a set of equal rank, disconnected nodes all of which are either heads or tails of arrows connecting them to nodes of other blocks. 
\end{definition}
Physically, this simply means that we have organised a set of gauge group factors, all of which are of equal rank and which have no bi-fundamental fields charged amongst them, into a ``block".
The whole set can then be described as a single node with a multiplicity denoting the cardinality of the set, and single arrows with multiplicities connecting this block to others. 
We see, indeed, that we are dis-allowing arrows which join nodes to themselves.
With this convention any chiral quiver diagram has a block structure with all blocks trivially having multiplicity one. We sketch these notions with an example in Fig.~\ref{fig:3b}. 
A block quiver can therefore be presented by the following data:
\begin{itemize}
\item the number of blocks,
\item the number of nodes in each block, and
\item the number of arrows connecting any pair of blocks.
\end{itemize}
Bearing this in mind, let us set the notation to be adopted in this work:
\begin{quote}
{\bf Notation }\label{notation}
Blocks are indexed by integers $i\in\{1,2,\ldots n\}$. 
We denote the number of nodes in block $i$ by $\alpha_i$ and the number of bi-fundamentals between blocks $i$ and $j$ with $a_{ij}$. The orientation of the arrows is taken into account by demanding $a_{ij}=-a_{ji}$. 
We encode the block structure of the quiver by writing its adjacency matrix as $q_n = \{a_{ij}\}$; clearly $q_n$ is a $n\times n$ antisymmetric matrix over $\mathbb{Z}$. 
Moreover, we let the R-charge of the bi-fundamental fields $a_{ij}$ be $r_{ij}$. 
Lastly, we write $N_i=N x_i$ for the rank of the gauge group of block $i$, with $N$ representing any common divisor of the ranks of all the blocks.
\end{quote}

Now, the main problem of our interest is the following,
\begin{quote}
{\sf Problem: }
{\it
Among all possible data $(N_i, \alpha_i, q_n)$ for block quivers, classify
those which may admit a consistent, superconformal quiver gauge theory in (3+1)-dimensions.
}
\end{quote}
The answer to this question, for $n=3$, has been given in both the mathematics and the physics literature \cite{1997alg.geom..3027K,Herzog:2003dj,Herzog:2003zc,Herzog:2004qw,Feng:2002kk,Benvenuti:2004dw}.

\subsection{Three-Block Quivers}\label{sec:3b}
We begin by reviewing the physics approach of \cite{Herzog:2003dj,Benvenuti:2004dw} for $n=3$.
Given a chiral quiver, like the one in the right of Fig.~\ref{fig:3b}, we must first clarify the 
necessary, though not sufficient, physical constraints that should be imposed in order to have a sensible superconformal gauge theory.

\paragraph*{Anomaly Cancellation: }
First, one has to make sure that the gauge (triangle ABJ) anomalies are cancelled. 
This is equivalent to the condition that the block-reduced rank vector $d = \{\alpha_i x_i\}$ of the quiver, lies in the kernel of the anti-symmetrized reduced quiver matrix $q_n$:
\begin{equation}\label{q3}
q_3\cdot d=0\;,\qquad \qquad
q_3 = 
\left(
\begin{array}{ccc}
0 & a_{12} & -a_{31} \\
-a_{12} & 0 & a_{23} \\
a_{31} & -a_{23} & 0\\
\end{array}
\right) \ ,
\quad
d :=
\left(\begin{array}{c}\alpha_1x_1\\\alpha_2x_2\\\alpha_3x_3\end{array}\right) \ ,
\end{equation} 
where the indices indicate the tail and the head of each arrow respectively. When the matrix indices are not in agreement with the arrow indices we write a minus sign.
Moreover, the quiver diagram must be free of {\it source} and {\it sink} configurations where source (sink) is a node with all incident arrows outgoing (incoming); this fixes an overall orientation of the quiver which we choose as counter-clockwise.
Now, the kernel of a $3\times3$ antisymmetric matrix is one dimensional, and for the given case the basis vector of ${\rm ker}(q_3)$ is simply
\begin{equation}\label{ker3}
d=\left(\begin{array}{c}a_{23}\\a_{31}\\a_{12}\end{array}\right) \ .
\end{equation}

\paragraph*{Beta functions: }
Next, we require the beta functions for each coupling present in the theory must vanish. 
The numerators of the beta functions are given by the $SU(N)$ NSVZ formula \cite{Novikov:1983uc}, 
which for the $i$th block reads
\begin{equation}\label{nsvz}
\beta_i=N_i + \sum_{A\in {\rm adj}[i]}N_i(r_{A,i}-1)+\frac{1}{2}\sum_{B\in {\rm bifund}[i,j]}N_j(r_{B,ij}-1) \ ,
\end{equation}
where $r$ is the R-charge of the fields, which are adjoints (adj) or bi-fundamentals (bifund).
Of course, we only have bi-fundamentals here.
Note that for our purpose, considering the numerators of the beta functions is enough, since the vanishing of the numerators is equivalent to the vanishing of the whole fraction, given that the denominator is finite.
 
Furthermore, in \cite{Henningson:1998gx}, it was shown that in a $(3+1)$-dimensional conformal field theory the gravitational central charges $c$ and $a$ are equal in the large $N$ limit, a result which was further generalized in \cite{Benvenuti:2004dw} to any superconformal quiver gauge theory. 
There, the authors used this fact to show that there is an extra condition on the beta functions
\begin{equation}\label{rel}
\lim_{N\rightarrow\infty}{\rm tr}R=\sum_i N_i\beta_i=c-a=0 \ .
\end{equation}

\paragraph*{Gamma functions (marginality): }
Conformality also requires the gamma functions to vanish. Our last physical input thus is the requirement that all the operators in the superpotential are marginal at the interacting superconformal fixed point, namely that they have R-charge equal to 2. 
The possible operators present in the superpotential for the three-block quiver on the right of Fig.~\ref{fig:3b} are its cyclic paths and correspond to cubic operators collectively represented as $$\hat{X}_{12}\hat{X}_{23}\hat{X}_{31} \ ,$$
with $\hat{X}_{ij}$ an arrow from block $i$ to block $j$. 
The marginality condition then translates to
\begin{equation}\label{3r}
r_{12}+r_{23}+r_{31}=2.
\end{equation}
Putting together the requirement \eqref{3r} and the vanishing of the beta functions \eqref{nsvz} for the three couplings results in a system of three unknown R-charges which satisfy four equations. Condition \eqref{rel} imposes the linear relation among the three beta functions that allows for a solution to the system. 
Substituting \eqref{ker3},\eqref{3r} in \eqref{rel} results in an equation in terms of the quiver data:
\begin{equation}\label{markov}
\frac{a_{23}^2}{\alpha_1}+\frac{a_{31}^2}{\alpha_2}+\frac{a_{12}^2}{\alpha_3}=a_{12}a_{23}a_{31} \ .
\end{equation}
This is a {\it Diophantine equation} in the variables $a_{ij}$ and $\alpha_i$, which are by definition integers.

For $\alpha_1=\alpha_2=\alpha_3=\alpha$, \eqref{markov} reduces to the well-studied {\em Markov equation}. 
This equation has solutions over $\mathbb{Z}$ which can be organized in a {\it tree} (cf.~\cite{Franco:2003ja}). 
This also holds for generic values of $\alpha_1,\alpha_2,\alpha_3$. More specifically, given a solution $(a_{23},a_{31},a_{12})$ one can construct an infinite set of solutions by the following operations:
\begin{equation}\label{markovtrans}
(a_{23},a_{31},a_{12})\rightarrow \Bigg \{ \begin{array}{c}(\alpha_1 a_{12}a_{31} - a_{23},a_{31},a_{12}) \\ (a_{23},\alpha_2 a_{12}a_{23}-a_{31},a_{12}) \\ (a_{23},a_{31},\alpha_3 a_{23}a_{31}-a_{12}) \end{array}\ .
\end{equation} 
In \cite{Feng:2000mi,Feng:2001bn,Franco:2003ja} it was shown how Seiberg duality can be represented as a quiver duality which can be described as follows: Pick a node to dualize, say node $k$; define three sets of arrows, $Q_{\rm in}, Q_{\rm out}$ and $Q_{\bar{k}}$ containing incoming, outgoing and non incident arrows with respect to the duality vertex; change the orientation of all arrows in $Q_{\rm in} \cup Q_{\rm out}$; change the arrows in $Q_{\bar{k}}$ as $a_{ij} \mapsto a_{ij} - a_{ik}a_{kj}$. Now recall that anomaly cancellation forces the rank of each node to be proportional to the number of its non incident arrows. This condition in combination with the operations described above, correctly reproduces the rank of the dualized node as $N_C^{\rm dual} = N_F-N_C$, where the number of flavors of the vertex $k$ is defined as
\begin{equation}\label{flavor}
N_F = \sum_{a_{jk} \in Q_{\rm in}} a_{jk} x_j = \sum_{a_{kj} \in Q_{\rm out}} a_{kj} x_j.
\end{equation} 
Note that the transformation \eqref{markovtrans} exactly matches the operations induced by Seiberg duality. Thus, the latter can be described as the action of the automorphism group on the Markov tree; we will return to this point on Seiberg duality in the next section.
In \cite{Benvenuti:2004dw} the solutions corresponding to the ``roots'' of the duality trees for generic values of the node multiplicities were found to be corresponding to all the three-block del Pezzo and pseudo del Pezzo quivers, as well as two new non-del Pezzo quivers which were dubbed ``shrunk", for it was shown that they arise from a specific operation (shrinking) on the block quiver. 

\subsection{The Markov Equation and the Adjacency Matrix}
Let us now derive the same Diophantine equation from a new perspective which does not require any physical input. As we shall see later, this form of the equation remains qualitatively the same for any number of blocks enabling us to extrapolate our results to such cases.
Recall that the $\{i,j\}$-th first minor $M_{ij}$ of an $n \times n$ matrix, is the determinant of the submatrix with row $i$ and column $j$ deleted while the $\{i,j\}$-th cofactor is given by $C_{ij} = (-)^{i+j}M_{ij}$. Because of the anti-symmetry of $q_3$, given in \eqref{q3}, one can see that its minors are quadratic in the edge multiplicities and the matrix of minors assumes the following simple form:
\begin{equation}\label{minor3}
M = \{M_{ij}\} = 
\left(
\begin{array}{ccc}
 a_{23}^2 & -a_{31} a_{23} & a_{12} a_{23} \\
 -a_{31} a_{23} & a_{31}^2 & -a_{12} a_{31} \\
 a_{12} a_{23} & -a_{12} a_{31} & a_{12}^2
\end{array}
\right)
\  .
\end{equation}
Now, the Diophantine equation \eqref{markov} can be written as a sum over the cofactors of the adjacency matrix, weighted by the respective elements of the quiver matrix and the block multiplicities $\alpha_i$. That is, equation \eqref{markov} can be represented as
\begin{equation} \label{comarkoff}
\sum_{i} C_{ii} \prod_{n \neq i}\alpha_n  - \sum_{i<j} q_3(i,j) C_{ij} \prod_n \alpha_n = 0 \ ,
\end{equation}
where the indices run in $\{1,2,3\}$.
The reason for writing the Markov equation in this form will become evident in the next subsection where we clarify its origin using representation theoretic concepts, while in Section~\ref{s:5b} we show that it is a general formula that applies for an arbitrary odd number of blocks and derive an analogous one for even numbers. 

Note that this construction, surprisingly suggests that all the necessary physical input of superconformality is somehow hidden in the adjacency matrix. For example, this formula requires neither the superpotential to be marginal nor the beta functions to vanish as conditions. The summation over minors automatically ensures these features! 

\subsection{The Markov Equation and the Tits Form}\label{3btits}
Equation \eqref{markov} has been derived in different contexts and via different routes. In the mathematics literature, using complete exceptional collections of coherent sheaves over del Pezzo surfaces \cite{1997alg.geom..3027K}, it was derived as a Diophantine equation which the ranks of the exceptional sheaves should satisfy\footnote{See \cite{Hanany:2006nm} for an equivalence between brane tilings and exceptional collections}. 
In \cite{Herzog:2003zc,Herzog:2004qw,Cecotti:2011rv,Alim:2011kw} these results were independently re-derived and linked with quiver gauge theories and Seiberg duality thereof, while in \cite{Cecotti:1992rm} this equation was derived using monodromy. 

In this subsection we will show how the generalized Markov equation in the form \eqref{comarkoff} is related to the Tits form of the quiver. For the sake of completeness we begin by briefly reviewing the basic facts about bilinear forms associated with quivers. A nice place where the interested reader can look for further background material on bilinear forms and the Tits form is \cite{ddpw} and  references therein.
\paragraph*{{\bf Bilinear Forms on Quivers}}
Given a quiver \qui$=(\mathbf{Q_0},\mathbf{Q_1})$, where $\mathbf{Q_0}$ denotes the set of vertices and $\mathbf{Q_1}$ the set of arrows, one can define a {\it representation} of \qui as the assignment of a vector space $V_i$ to each vertex $i\in\mathbf{Q}_0$ and a linear map $V_\rho:\;V_{t_{(\rho)}}\mapsto V_{h_{(\rho)}}$ to each arrow $\rho\in\mathbf{Q}_1$, with the subscripts $t$ and $h$ denoting the tail and the head of an arrow respectively. We call the vector $\{{\rm dim}V_i\}$ the {\it representation vector} of the quiver. A {\it path algebra}, with the product operation given by concatenation of arrows, can be associated to a quiver. In the cases that we consider here there is also a set of algebraic relations $\mathcal{F}$, that the arrows obey, which come from the F-terms (superpotential) of the gauge theory. Quivers with such relations are called {\it bounded}. The quotient of the path algebra by $\mathcal{F}$ yields the so-called  
{\it $\cal{F}$-flat} or {\it Jacobian algebra} $A$ of \qui. 
%

On the modules of $A$, we can define
the {\it Euler form} as the bilinear form given by

\begin{equation}\label{euler-form}
\langle \mathbf{x},\mathbf{y} \rangle = \sum_{i\in\mathbf{Q_0}}x_iy_i-\sum_{\rho\in\mathbf{Q_1}}x_{t(\rho)}y_{h(\rho)}
\ ,
\end{equation}
where $\mathbf{x} \equiv \{{\rm dim}V_i\}$.
The symmetrization of \eqref{euler-form}, referred to as the {\it Cartan form}, can be written as
\begin{equation} \label{cartan-form}
(\mathbf{x},\mathbf{y})=\langle \mathbf{x},\mathbf{y} \rangle+\langle \mathbf{y},\mathbf{x} \rangle=\mathbf{x}^TC_Q\mathbf{y} \ ,
\end{equation}
where $C_Q=(c_{i,j})_{i,j\in\mathbf{Q_0}}$ is a symmetric $|\mathbf{Q_0}|\times|\mathbf{Q_0}|$ generalized Cartan matrix with $\mathbb{Z}$ valued entries, given by
\begin{equation}\label{cij}
c_{i,j}=\Bigg\{\begin{array}{cc} 2-2\#(\text{loops at $i$}),& \text{if}\quad i=j\\
-\#\text{arrows between $i$ and $j$},&\text{if}\quad i\neq j
\end{array}
\end{equation}
%
Last, we define the {\it Tits form} which is the quadratic form associated with the Euler form,
\begin{equation}\label{tits-form}
q_{\mathbf Q}(\mathbf{x})\equiv(\mathbf{x},\mathbf{x})=\sum_{i\in\mathbf{Q_0}}x_i^2-\sum_{\rho\in\mathbf{Q_1}}x_{t(\rho)}x_{h(\rho)}=\frac{1}{2}\mathbf{x}^TC_Q\mathbf{x}
\end{equation}

We are now in a position to relate these concepts to the block quivers. 
The Tits form for the three-block case of Fig.~\ref{fig:3b} using the adjacency matrix \eqref{q3} reads:
\begin{equation}\label{tits3b}
\begin{split}
q_Q(\mathbf{x}) & = q_Q(x_1, x_2, x_3)
\equiv\sum_{i\in{\mathbf Q_0}}x_i^2-\sum_{i<j}|q_3(i,j)|x_ix_j \\ & = 
\alpha_1 x_1^2 + \alpha_2 x_2^2 + \alpha_3 x_3^2 - a_{12}x_1 x_2 \alpha_1\alpha_2- a_{23}x_2 x_3 \alpha_2\alpha_3 - a_{31}x_1 x_3 \alpha_1\alpha_3
\end{split}
\end{equation}
The Markov equation though is given by a slightly modified form. For that, let us consider an orientation dependent version of \eqref{tits3b}, which we call $q_{Q_s}$ ($s$ for ``signed''), without the absolute value in the adjacency matrix elements. As we immediately show this form yields the desired Diophantine equation whose roots label superconformal block quivers. We then connect $q_{Q_s}$ with the Tits form $q_{Q}$. For the three-block case it reads:
\begin{equation}\label{tits3br}
\begin{split}
q_{Q_s}(\mathbf{x}) & = q_{Q_s}(x_1, x_2, x_3)
\equiv\sum_{i\in{\mathbf Q_0}}x_i^2-\sum_{i<j}q_3(i,j)x_ix_j \\ & = 
\alpha_1 x_1^2 + \alpha_2 x_2^2 + \alpha_3 x_3^2 - a_{12}x_1 x_2 \alpha_1\alpha_2- a_{23}x_2 x_3 \alpha_2\alpha_3 + a_{31}x_1 x_3 \alpha_1\alpha_3 .
\end{split}
\end{equation}
After setting 
\begin{equation}\label{titstomarkov3b}
x_1 = \sqrt{\frac{\alpha_2\alpha_3}{\alpha_1 K^2}}a_{23} \; , \; x_2 = \sqrt{\frac{\alpha_1\alpha_3}{\alpha_2 K^2}}a_{31} \; , \; x_3 = \sqrt{\frac{\alpha_1\alpha_2}{\alpha_3 K^2}}a_{12},
\end{equation}where $K^2=12(9)-(\alpha_1 + \alpha_2 + \alpha_3)$ for a del Pezzo(non del Pezzo) quiver \cite{1997alg.geom..3027K,Benvenuti:2004dw} ensures that $x_i \in \mathbb{Z}$, we are left precisely with the Markov equation \eqref{markov}! These conditions are exactly those found in \cite{1997alg.geom..3027K} (cf. Sec 3) in the context of exceptional collections of sheaves and coincide with the anomaly cancellation \eqref{ker3}.
Going back to the matrix of minors \eqref{minor3}, we immediately see that setting $q_{Q_s}=0$ yields the minor summation formula \eqref{comarkoff} with $C_{ij} = x_i x_j$.

%
Now, by adding and subtracting the term $a_{31}x_1 x_3 \alpha_1\alpha_3$ to \eqref{tits3b}, we obtain $q_{Q} = q_{Q_s} - 2a_{31}x_1 x_3 \alpha_1\alpha_3. $ Using the fact that we are looking for solutions of the Markov equation, i.e. $q_{Q_s} = 0,$ we see that the dimension vector of a quiver gauge theory of this class satisfies
\begin{equation}\label{tits3bf}
q_Q(x_1, x_2, x_3) = - 2 |q(3,1)|\alpha_1\alpha_3 x_1 x_3 \ .
\end{equation}
Using the relations \eqref{titstomarkov3b} this equation can be rewritten as
\begin{equation}\label{tits3bf-gen}
q_Q(x_1, x_2, x_3) = - 2 \sqrt{\alpha_1 \alpha_2 \alpha_3 K^2} x_1 x_2 x_3 \ .
\end{equation}
with $\sqrt{\alpha_1 \alpha_2 \alpha_3 K^2}\in\mathbb{Z}$.
Thus, we arrive at the conclusion that {\it the dimension vectors of superconformal block quiver theories have a negative Tits form}\footnote{See \cite{Cecotti:2012va} for the appearance of the Tits form in ${\cal N}=2$ quivers.}. 

The Tits form is important because it defines the {\it type} of a quiver: positive definite, positive semi-definite and indefinite correspond respectively to finite, tame and wild types (cf.~\cite{Savage}).
Moreover, together with its extension given by Kac, the Tits form provides the link between quiver representations and root systems. 

We can associate the dimension vector (i.e., the vector whose entries are the ranks of the gauge group factors) to a root of the root system of the underlying quiver and the Cartan form \eqref{cartan-form} to the inner product on the root space. The Tits form is therefore the norm-squared of a root vector. 
It is a celebrated theorem of Kac (see App.~\ref{ap:quiverrjpg}) that real roots correspond to quivers with exactly one indecomposable representation and the norm-squared of the dimension vector is equal to 1; in contrast, imaginary roots correspond to the case where there are families of indecomposable representations and the norm-squared is less than or equal to 0.

In light of Kac's theorem the fact that we have a negative norm means that our choices of dimension vectors, imposed by superconformality, correspond to imaginary roots of the root system associated with the quiver. 
The ranks of superconformal gauge theories form the subset of such dimension vectors that  satisfy \eqref{tits3bf}.
It would be very interesting to see if these physically special quiver representations also have special algebraic properties\footnote{An example of a root with a special property is a so called {\it Schur} root, which corresponds to a dimension vector of an indecomposable representation \cite{Derksen}.} which could shed some light in the study of wild quivers. Since very little is known on that subject we will not try to address this question here, but will leave it as an interesting comment.
\subsection{Seiberg Duality and the Affine Weyl Group}\label{3bseiberg}
In \cite{Cachazo:2001sg} Seiberg duality was interpreted as the action of the affine Weyl group on the root system of an (affine) A-D-E type quiver diagram.
As we now show, in our construction this result can be generalized to arbitrary three-block quivers. We will later see that this statement actually holds for any odd block number. Before discussing that let us briefly remind the reader how one defines the {\it Weyl group} of the root system associated to a quiver \qui and connects it with the classification scheme of finite-tame-wild. The idea behind this construction is to think of the vector space spanned by the dimension vectors of the quiver as a root space of the algebra associated with the quiver.

For simplicity we write the vertex set as ${\mathbf Q_0}=\{1,2,...,n\}$ and denote the corresponding basis of $\mathbb{Z}\mathbf{Q_0}$ as $\mathbf{e_1},...,\mathbf{e_n}.$ For each vertex $i\in\mathbf{Q_0}$ define an element $r_i\in{\rm Aut}(\mathbb{Z}\mathbf{Q_0})$ whose action on a dimension vector $\mathbf{x} \in \mathbb{Z}\mathbf{Q_0}$ reads
\begin{equation} \label{weyl-action-1}
r_i[\mathbf{x} ]=\mathbf{x}  - 2\frac{(\mathbf{x} ,\mathbf{e_i})}{(\mathbf{e_i} ,\mathbf{e_i})}\mathbf{e_i} = \mathbf{x}  - (\mathbf{x} ,\mathbf{e_i})\mathbf{e_i} 
\end{equation}where the inner product $(-,-)$ is given by the Cartan form \eqref{cartan-form}.
%
%
%
If there are no loops at vertex $i$ then we call $r_i$ a simple reflection and $\mathbf{e_i}$ a simple root. One can easily check that a simple reflection leaves the Tits form \eqref{tits-form} invariant. 
The Weyl group $W(Q)$ of the quiver is defined as the subgroup of ${\rm Aut}(\mathbb{Z}\mathbf{Q_0})$ generated by the simple reflections $r_i$. 

Let us now adapt this discussion in the three-block quivers depicted in Fig.~\ref{fig:3b} and see how Seiberg duality arises in this context. Note that since we allow for arbitrary number of arrows between two nodes, we are not restricted to an A-D-E quiver diagram. To illustrate the idea with a simple example, we first focus in the case where all block multiplicities $\alpha_i$ are set to one, and we will then generalize to arbitrary numbers. By writing the Tits form \eqref{tits3b} as $$(\mathbf{x},\mathbf{x}) = \frac{1}{2}\mathbf{x}^TC_Q\mathbf{x}$$
one can read off the Cartan matrix of a three-block quiver. That is:
\begin{equation}\label{cartan3b}
C_Q=\left(\begin{array}{ccc}2 & -a_{12} & -a_{31} \\  -a_{12} & 2 & -a_{23} \\ -a_{31} & -a_{23} & 2 \end{array}\right)
\end{equation} 
In general, one can define a Cartan matrix as $C = 2\mathbb{I} - q$, where $q$ is the adjacency matrix defined irrespectively of the orientation of the arrows. In our case we have defined $q_3$ as the antisymmetrized adjacency matrix \eqref{q3}, hence this relation does not hold. The Cartan matrix is symmetric, so it is associated to a simply-laced algebra, and it can be easily shown that it is indefinite. That is it has both positive and negative principal minors. It thus describes some Kac-Moody algebra of {\it indefinite} type, in accordance with the fact that we are dealing with wild quivers. 

To proceed, consider the reflection of a vector $\mathbf{x} = (x_1, x_2, x_3)^{\rm T}$ with respect to the simple root $\mathbf{e_1} = (1,0,0)^{\rm T}$. We have
\begin{equation}\label{weyl-seiberg-1}
r_1[\mathbf{x}] = \mathbf{x} - (2 x_1 - a_{12} x_2 - a_{31} x_3) \mathbf{e_1} = \left(\begin{array}{c} a_{12} x_2 +  a_{31} x_3 - x_1  \\ x_2 \\ x_3 \end{array}\right).
\end{equation}
Recall that $\mathbf{x}$ is the dimension vector representing a superconformal gauge theory and Seiberg duality is described as the transformation \eqref{markovtrans} and the operations outlined in the paragraph right below it. In addition, the rank of the node with label ``1'' is $x_1 = N_{C_1}$, the number of flavors is given by \eqref{flavor} as $N_{F_1} = a_{12} x_2 = a_{31} x_3$ while the rank of the dualized node reads $N_C^{dual} = N_F - N_C$. We thus see that the top component of the right hand side of \eqref{weyl-seiberg-1} is the rank of the first node when Seiberg dualized, plus a shift. Therefore, we arrive at the following realization of Seiberg duality in terms of roots:
%
\begin{equation}\label{seiberg-reflection}
S_i[\mathbf{x}] = r_i[\mathbf{x}] - N_{F_i}\mathbf{e_i} , 
\end{equation}
where $S_i[\mathbf{x}] $ denotes Seiberg duality of the quiver gauge theory $\mathbf{x}$ with respect to node ``i''. Such an operation is known as an {\it affine reflection}.
As we now show, affine reflections leave the Markov equation invariant. Let us demonstrate that by computing the Diophantine equation for the dualized quiver with block multiplicities equal to one. Recall that the Markov equation can be written (cf. \eqref{tits3bf}) in the form
$$
(\mathbf{x}, \mathbf{x}) = - 2 \sqrt{K^2} \prod_j x_j.
$$
Using \eqref{seiberg-reflection} we find that the norm-squared of a vector dualized with respect to block ``$i$'' reads
\begin{equation}\label{seiberg-closure}
(S_i[\mathbf{x}] , S_i[\mathbf{x}]) = - 2 \sqrt{K^2} x_i' \prod_{j\neq i} x_j ,
\end{equation}
where $x_i' = a_{ij} x_j - x_i = N_{F_i} - N_{C_i}$. 
This nicely demonstrates that the subset of roots that correspond to superconformal gauge theories is closed under Seiberg duality. In other words, this results asserts that superconformal gauge theories are special roots of the quiver algebra and Seiberg duality corresponds to the action of the affine Weyl group on the root system.

We now repeat the discussion for generic three-block quivers. Had we followed the same method as right above we would have ended up with a Cartan matrix of the form
\begin{equation}\label{cartan3b-fake}
C_Q=\left(\begin{array}{ccc}2\alpha_1 & -\alpha_1 \alpha_2 a_{12} & -\alpha_1 \alpha_3 a_{31} \\  -\alpha_1 \alpha_2 a_{12} & 2 \alpha_2 & -\alpha_2 \alpha_3 a_{23} \\ -\alpha_1 \alpha_3 a_{31} & -\alpha_2 \alpha_3 a_{23} & 2 \alpha_3 \end{array}\right).
\end{equation} 
Recall that a Cartan matrix should have 2's in the diagonal. The $\alpha_i$ factors in \eqref{cartan3b-fake} are due to the block reduction of the quiver. In order to construct the correct $C_Q$, we should instead consider the nodes in each block as independent entries in the adjacency matrix. By doing that we obtain a matrix of dimension $\sum_i \alpha_i \times \sum_i \alpha_i$ with the desired property. In other words we consider the Tits form as $(\sum \alpha_i)$-ary quadratic form, where the first $\alpha_1$ variables degenerate to $x_1$, the second $\alpha_2$ to $x_2$ and the last $\alpha_3$ to $x_3$. We therefore have

\begin{equation}\label{cartan-eb-general}
 C_Q = \;\; \begin{blockarray}{ccccccccc}
     \begin{block}{(ccc|ccc|ccc)}
       & & & -a_{12} & \ldots &-a_{12} & -a_{13} & \ldots & -a_{13} \\
       & 2\mathbb{I}_{\alpha_1} &  & \vdots & \ddots & \vdots & \vdots & \ddots & \vdots \\
       & & & -a_{12} &\ldots &-a_{12} & -a_{13} & \ldots & -a_{13} \\
       \cline{1-9}
  \begin{block*}{(ccc|ccc|ccc)}
   -a_{12} & \ldots &-a_{12} & & & & -a_{23} & \ldots & -a_{23} \\
    \vdots & \ddots & \vdots & & 2\mathbb{I}_{\alpha_2} & & \vdots & \ddots & \vdots  \\
    -a_{12} & \ldots &-a_{12} & & & & -a_{23} & \ldots & -a_{23} \\
    \cline{1-9}
  \begin{block*}{(ccc|ccc|ccc)}
  -a_{13} & \ldots &-a_{13} & -a_{23} & \ldots & -a_{23} & & & \\
    \vdots & \ddots & \vdots & \vdots & \ddots & \vdots & & 2\mathbb{I}_{\alpha_3} &  \\
    -a_{13} & \ldots &-a_{13} & -a_{23} & \ldots & -a_{23} & & & \\
  \end{block*}  
  \end{block*}
  \end{block}
  \end{blockarray}\;\;\;\;,
\end{equation}

The basis vector, with respect to which we are going to reflect, is ${\mathbf a} = (0,\dots,0,\underbrace{1,\dots,1}_{\alpha_{i}},0,\dots,0)$ with ones in the $i$-th $\alpha_i$ entries and zeros in the rest. This corresponds to duality of block $i$. The norm of this root is given by $({\mathbf a},{\mathbf a})= 2\alpha_i$. The inner product of a generic vector $\mathbf{x}=(\underbrace{x_1,\dots,x_1}_{\alpha_{1}},\underbrace{x_2,\dots,x_2}_{\alpha_{2}},\underbrace{x_3,\dots,x_3}_{\alpha_{3}})$ with ${\mathbf a}$ is given by 
$$
({\mathbf x},{\mathbf a}) = \alpha_i (2x_i - a_{ij}\alpha_j x_j - a_{ik}\alpha_k x_k)
$$
where $j,k$ index the other two blocks. Using the definition of the flavor number, this expression reads 
\begin{equation}
({\mathbf x},{\mathbf a}) = 2\alpha_i (N_{C_i} - N_{F_i})
\end{equation}
Using the reflection formula \eqref{weyl-action-1} we see that \eqref{seiberg-reflection} generalizes to
\begin{equation}\label{seiberg-reflection-gen}
S_i[\mathbf{x}] = r_i[\mathbf{x}] - N_{F_i}\mathbf{a} , 
\end{equation}
and computing the norm of the dualized vector, we find that it obeys the relation
\begin{equation}\label{norms-gen}
(S_i[\mathbf{x}] , S_i[\mathbf{x}]) = (\mathbf{x} , \mathbf{x}) + 2 \alpha_i N_{F_i}(N_{C_i} - N_{C_i}').
\end{equation}
%
Using \eqref{tits3bf-gen} we arrive at the following result
\begin{equation}\label{seiberg-closure-gen}
(S_i[\mathbf{x}] , S_i[\mathbf{x}]) = - 2 \sqrt{\alpha_1 \alpha_2 \alpha_3 K^2} x_i' \prod_{j\neq i} x_j.
\end{equation}
%
That is, Seiberg duality corresponds to an affine Weyl reflection for any three-block quiver, where duality with respect to block ``$i$'' maps to reflection with respect to the vector ${\mathbf a}$ with ones in the $i$-th $\alpha_i$ entries and zeros in the rest. The form of the Diophantine equation remains invariant under this operation so that if $\mathbf{x}$ is solution, another one can be obtained as $S_i[\mathbf{x}]$. 

\paragraph*{Summary: }

In this section we reviewed the concept of block quivers and using new techniques, 
we re-derived results well-known in both the mathematics and the physics literature for
the case of three-block quivers.
We have shown that conformality of the gauge theory, 
which the chiral quiver encodes, and anomaly cancellation place constraints on the 
adjacency and rank data of the quiver, in the form of a Diophantine 
equation which can be presented as a weighted sum over minors of the adjacency matrix.
For three-blocks, this is a (generalized) Markov equation.

We then recalled standard techniques of representation theory of quivers.
In particular, we used the Tits bilinear form defined on the space of 
dimension vectors - or root space - of the quiver.
We showed that a signed version of the Tits form is precisely the 
aforementioned Diophantine equation justifying the minor formula. This allowed for a correspondence
between superconformal gauge theories and root vectors of the quiver's root system.
In the ensuing section, we will see that our results persist for an 
arbitrary block quiver.

Finally, on quiver theories in our context, there is the famous Seiberg 
duality action.
We saw that this translates to an affine Weyl reflection on the root 
space under which the Diophantine equation
remains invariant. This is in accord with the fact the duality tree of 
Seiberg-dual theories are classified by solutions of our Diophatine 
equation~\cite{Franco:2003ja}.

\section{New Results for Higher Block Number}\label{s:5b}\setall
\comment{Having seen the nice story under our new light for the three-block case,} Having reviewed the three-block case under a new perspective, one naturally wonders how to proceed to higher number of blocks. In this section we will generalize our previous discussion to four- and five-block quivers and then conjecture the form of the superconformality conditions for any number of blocks.

\subsection{Four-Block Models}
Now, the four-block situation was also addressed in \cite{Benvenuti:2004dw} and we refer the reader to the classification therein.
We remark that for $n=4$ there is a unique choice to draw a quiver with no sink or source configurations. Furthermore, since we have an antisymmetric matrix of even dimension as the reduced adjacency matrix, the determinant does not vanish automatically and the situation is a little more difficult
to regard it fully in terms of our quadratic form analysis. We are able though to unravel a similar structure as a sum of minors for the four block case as well. The adjacency matrix that we will consider is
\begin{equation}\label{q4}
q_4 = 
\left(
\begin{array}{cccc}
0 & a_{12} & -a_{13} & -a_{14} \\
-a_{12} & 0 & a_{23} & -a_{24} \\
a_{13} & -a_{23} & 0 &  a_{34} \\
a_{14} & a_{24} & - a_{34} &  0
\end{array}
\right),
\end{equation}
such that \begin{equation}\label{detq4}\det q_4= a_{41} a_{23} + a_{31} a_{42} - a_{12} a_{34} = 0,\end{equation} as required for anomaly cancellation. The $3\times 3$ first minors of the adjacency matrix vanish since they are proportional to $\det q_4$. Therefore let us consider the $2\times 2$ second minors of \eqref{q4} where a second minor $M_{ij,kl}$ is defined as the determinant of the submatrix that results if one removes the $i,j$ rows and the $k,l$ columns of the original matrix. The relevant minor matrix for our case is
\begin{equation}
M_{q_4}= \begin{blockarray}{ccccccc}
    & 12 & 13 & 14 & 23 & 24 & 34  \\ 
\begin{block}{c\Left{}{(\mkern1mu}cccccc<{\mkern1mu})}
 12 \;\;\;\;& a_{34}^2 & -a_{24} a_{34} & a_{23} a_{34} & -a_{14} a_{34} & -a_{13} a_{34} & a_{12} a_{34} \\
 13 \;\;\;\;& -a_{24} a_{34} & a_{24}^2 & -a_{23} a_{24} & a_{14} a_{24} & a_{13} a_{24} & -a_{12} a_{24} \\
 14 \;\;\;\;& a_{23} a_{34} & -a_{23} a_{24} & a_{23}^2 & -a_{14} a_{23} & -a_{13} a_{23} & a_{12} a_{23} \\
 23 \;\;\;\;& -a_{14} a_{34} & a_{14} a_{24} & -a_{14} a_{23} & a_{14}^2 & a_{13} a_{14} & -a_{12} a_{14} \\
 24 \;\;\;\;&-a_{13} a_{34} & a_{13} a_{24} & -a_{13} a_{23} & a_{13} a_{14} & a_{13}^2 & -a_{12} a_{13} \\
 34 \;\;\;\;& a_{12} a_{34} & -a_{12} a_{24} & a_{12} a_{23} & -a_{12} a_{14} & -a_{12} a_{13} & a_{12}^2 \\
 \end{block}
\end{blockarray} \;\;\; ,
\end{equation}
where the outer column and row indicate the set of rows and columns of the adjacency matrix that are deleted in order to obtain the corresponding element of the minor matrix, e.g. $M_{q_4}(2,3) \equiv M_{13,14} = -a_{23}a_{24}$.
The Diophantine equation whose solutions are in one to one correspondence with the superconformal four-block quivers can be written as
\begin{small}
\begin{equation}\label{4b-form}
\sum_{i<j} M_{ij,ij} \prod_{m\neq i,j}\alpha_m - \sum_{i\neq j<k} (-)^{j+k} q_4(j,k) M_{ij,ik} \prod_{m\neq i}\alpha_m + \sum_{i<j<k<l} q_4(i,j)q_4(i,l) M_{ij,il} \prod_{m}\alpha_m = 0
\end{equation}
\end{small}
supplemented by \eqref{detq4}, where $\alpha_i$ denotes the multiplicity of the $i$-th block and the indices run in $\{1,2,3,4\}$. By substituting the minors one recovers the Diophantine equation reported in \cite{Benvenuti:2004dw},
\begin{equation}\label{4b-ami}
\begin{split}
\frac{a_{12}^2}{\alpha_3 \alpha_4} + \frac{a_{13}^2}{\alpha_2 \alpha_4} + \frac{a_{14}^2}{\alpha_2 \alpha_3} + \frac{a_{23}^2}{\alpha_1 \alpha_4} & + \frac{a_{24}^2}{\alpha_1 \alpha_3} + \frac{a_{34}^2}{\alpha_1 \alpha_2} + \frac{a_{12} a_{24} a_{14}}{\alpha_3} - \frac{a_{12} a_{23} a_{13}}{\alpha_4} \\ & + \frac{a_{13} a_{34} a_{14}}{\alpha_2} - \frac{a_{23} a_{34} a_{24} }{\alpha_1} - a_{12} a_{23} a_{34} a_{14} = 0
\end{split}
\end{equation}
Although this formula seems somehow arbitrarily written there is a check for its validity and that is the way it reduces to the three-block equation. Let us see what happens when we remove the block with label one for example.
This corresponds to the deletion of the first column and first row of the adjacency matrix $q_4$, leaving us with a three-block model adjacency matrix identical to $q_3$ in \eqref{q3}, while we also set $\alpha_1$ to zero. From the sum \eqref{4b-form} we see that the only terms remaining are the ones that are not multiplied by $\alpha_1$. These are,
\begin{eqnarray}\label{4b-to-3b}
 M_{12,12}\alpha_3\alpha_4 &+& M_{13,13}\alpha_2\alpha_4 + M_{14,14}\alpha_2 \alpha_3 \nonumber\\  &+& \Big( q_4(2,3) M_{12,13} - q_4(2,4) M_{12,14} + q_4(3,4) M_{13,14} \Big) \alpha_2 \alpha_3 \alpha_4  = 0
\end{eqnarray}
Now the remaining elements of $q_4$ become entries of $q_3$ as $q_4(i,j) \mapsto q_3(i-1,j-1)$ while the $2\times 2$ minors of $q_4$ become the first minors of $q_3$ and together with the sign $(-)^{j+k}$ yield the cofactors of $q_3$ as $(-)^{j+k}M^{(q_4)}_{ij,ik} \mapsto C_{j-1,k-1}^{(q_3)}$. Having this relation in mind one can immediately see that the formula \eqref{4b-form} correctly reduces to \eqref{comarkoff}! In the next paragraph we will see that the five-block Diophantine equation is identical to the three-block one. This statement, in combination with the fact that the reduction from four to three blocks can be demonstrated using the minor sum implies, inductively, that the formula \eqref{4b-form} holds also for six-block quivers. Thus, one can justifiably extrapolate this claim to any even number of blocks.

In this case though the relation with a bilinear form on the quiver is not clear. Since the summation over minors suggests a continuation from the three blocks, it is natural to think that an analogous perspective would be valid for the four blocks too. We leave this investigation for future work.

\subsection{Five-Block Models}
Let us move on to the next case of $n=5$.
We are looking for quivers with five blocks where there are no sink or source configurations. 
We will readily see that we now encounter a new situation.
For $n=3,4$, the possible topologies of such graphs were unique,
but this is not the case for order five and higher. 
Hence one has to count all such connected sinkless-sourceless graphs and mod out by topological equivalence, where we consider two graphs equivalent if they are related by a permutation of the edges and nodes.

\subsubsection{The Inequivalent Graphs}
We find six equivalence classes, the representatives of which we refer to as Type I to VI. We draw them in Fig.~\ref{fig:6types} and we also list the oriented cycles which correspond to operators in the superpotential. The cycle structure of the six types is summarized as follows (outdegree refers to the number of arrows going out of the node):

\begin{figure}[tbh!!!]
\begin{center}
\includegraphics[scale=0.3]{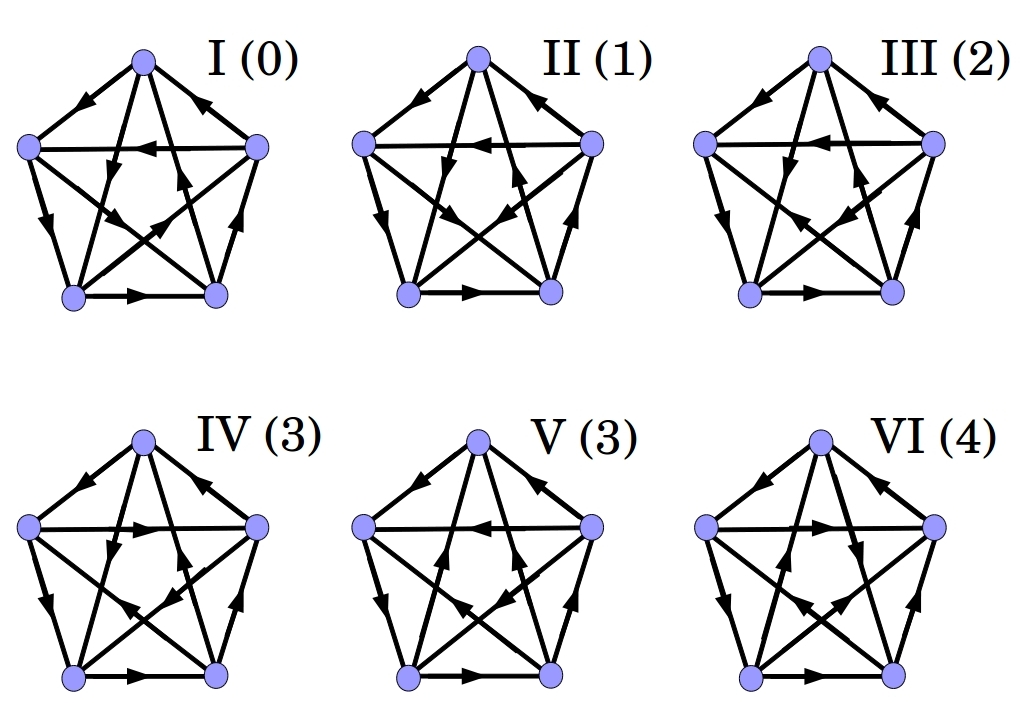}   
\end{center}
\caption{{\sf The six inequivalent chiral five-block quivers. The numbers in brackets indicate the number of clockwise internal (the ones not in the perimeter of the pentagon) arrows.}}
\label{fig:6types}
\end{figure}
\newpage
\paragraph*{Cycle counting}
\begin{itemize}
\item Type I: clockwise outdegrees (starting from mid top) $(2,2,2,2,2)$; 12 cycles; 2 quintics, 5 quartics, 5 cubics
\item Type II: clockwise outdegrees $(2,3,2,1,2)$; 9 cycles; 1 quintic, 4 quartics, 4 cubics
\item Type III: clockwise outdegrees $(2,3,3,1,1)$; 7 cycles; 1 quintic, 3 quartics, 3 cubics
\item Type IV: clockwise outdegrees $(2,2,3,1,2)$; 10 cycles; 3 quintics, 3 quartics, 4 cubics
\item Type V: clockwise outdegrees $(1,3,3,2,1)$; 6 cycles; 1 quintic, 2 quartics, 3 cubics
\item Type VI: clockwise outdegrees $(2,1,2,3,2)$; 9 cycles; 2 quintics, 3 quartics, 4 cubics
\end{itemize}

\comment{We also remind the reader of our notation that nodes are labelled from 1 to 5, with te rank being $N_i$, the multiplicity of the arrow from vertex $i$ to vertex $j$ is $a_{ij}$, its r-charge is $r_{ij}$.  The multiplicity of node $i$ is $\alpha_i,\beta_i,\gamma_i,\delta_i,\jpgilon_i,\eta_i$ for Type I,II,III,IV,V,VI respectively.}

\subsubsection{Detailed Analysis of Type I}\label{t1}
\begin{figure}[tbh]
\begin{center}
\includegraphics[scale=0.25]{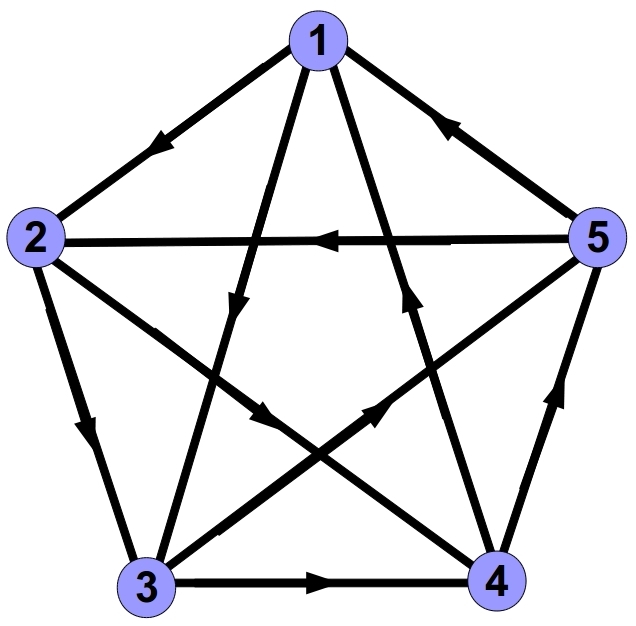}   
\end{center}
\caption{{\sf The quiver for Type I of the five-block theory.}}
\label{fig:type1}
\end{figure}

Let us begin with a detailed analysis of Type I, whose block quiver is given in Figure \ref{fig:type1}.

As in the three-block case we are going to impose the following conditions: 
\begin{enumerate}
\item anomaly cancellation: the dimension vector lies in the kernel of the quiver reduced adjacency matrix $q_5$;
\item beta functions: the weighted sum of the beta functions vanish;
\item gamma functions: R-charge of each cycle sums to 2.
\end{enumerate}
Now, following our previous notation, the adjacency matrix is
\begin{equation}\label{q5}
q_5=\left(\begin{array}{ccccc}
 0 & a_{12} & a_{13} & -a_{41} & -a_{51} \\
 -a_{12} & 0 & a_{23} & a_{24} & -a_{52} \\
 -a_{13} & -a_{23} & 0 & a_{34} & a_{35} \\
 a_{41} & -a_{24} & -a_{34} & 0 & a_{45} \\
 a_{51} & a_{52} & -a_{35} & -a_{45} & 0 
\end{array}\right) .
\end{equation}
The first condition then reads (recall that the rank is $N_i = N x_i$):
\begin{equation}
q_5 \cdot (\alpha_1 x_1,\alpha_2 x_2,\alpha_3 x_3, \alpha_4 x_4, \alpha_5 x_5)^\top={\mathbf 0}
\ ,
\end{equation} 
which translates to
\begin{eqnarray}\label{kernel1}
\alpha_1 x_1 & \propto & a_{45}a_{23} - a_{35}a_{24} - a_{52}a_{34} \equiv A_1 \nonumber\\
\alpha_2 x_2 & \propto & a_{51}a_{34} - a_{45}a_{13} - a_{41}a_{35} \equiv A_2 \nonumber\\
\alpha_3 x_3 & \propto & a_{45}a_{12} - a_{51}a_{24} - a_{41}a_{52} \equiv A_3 \\
\alpha_4 x_4 & \propto & a_{23}a_{51} - a_{35}a_{12} - a_{13}a_{52} \equiv A_4 \nonumber\\
\alpha_5 x_5 & \propto & a_{34}a_{12} - a_{41}a_{23} - a_{13}a_{24} \equiv A_5 \nonumber .
\end{eqnarray}
These equations can be nicely summarized as
\begin{equation}
\alpha_{i_1} x_{i_1} \propto \frac{1}{8}\epsilon_{i_1i_2i_3i_4i_5}a_{i_2i_3}a_{i_4i_5},
\end{equation} 
where summation over repeated indices is implied.
Next, using the NSVZ numerators for the beta functions, we find
\begin{equation}
\begin{split}
\beta_1 \!& = \! Nx_1\! +\! \frac{N}{2}\Big(\! A_5a_{51}(r_{51} \!-\! 1) \!+\! A_4a_{41}(r_{41}\!-\!1) \!+\! A_3a_{13}(r_{13}\!-\!1) \!+\! A_2a_{12}(r_{12} \!-\! 1)\!\Big) \\
\beta_2 \!& = \! Nx_2\! +\! \frac{N}{2}\Big(\! A_1a_{12}(r_{12} \!-\! 1) \!+\!A_5a_{52}(r_{52}\!-\!1) \!+\! A_4a_{24}(r_{24}\!-\!1) \!+\! A_3a_{23}(r_{23} \!-\! 1)\!\Big) \\
\beta_3 \!& = \! Nx_3\! +\! \frac{N}{2}\Big(\! A_2a_{23}(r_{23} \!-\! 1) \!+\! A_1a_{13}(r_{13}\!-\!1) \!+\! A_5a_{35}(r_{35}\!-\!1) \!+\! A_4a_{34}(r_{34} \!-\! 1)\!\Big) \\
\beta_4 \!& = \! Nx_4\! +\! \frac{N}{2}\Big(\! A_3a_{34}(r_{34} \!-\! 1) \!+\! A_2a_{24}(r_{24}\!-\!1) \!+\! A_1a_{41}(r_{41}\!-\!1) \!+\! A_5a_{45}(r_{45} \!-\! 1)\!\Big) \\
\beta_5 \!& = \! Nx_5\! +\! \frac{N}{2}\Big(\! A_1a_{51}(r_{51} \!-\! 1) \!+\! A_2a_{52}(r_{52}\!-\!1) \!+\! A_3a_{35}(r_{35}\!-\!1) \!+\! A_4a_{45}(r_{45} \!-\! 1)\!\Big) .
\end{split}
\end{equation}
Finally, since our graph has twelve oriented cycles which could contribute to the superpotential, we have twelve equations that the R-charges of the various operators should satisfy in order to have total R-charge equal to two for each cycle. 
These are:
\begin{eqnarray}\label{relations1}
\label{qui1} r_{12} + r_{23} + r_{34} + r_{45} + r_{15} & = & 2 \\
\label{qui2} r_{13} + r_{14} + r_{24} + r_{25} + r_{35} & = & 2 \\  
\label{qua1} r_{15} + r_{35} + r_{23} + r_{12} & = & 2 \\
\label{qua2} r_{15} + r_{45} + r_{34} + r_{13} & = & 2 \\
\label{qua3} r_{15} + r_{45} + r_{24} + r_{12} & = & 2 \\
\label{qua4} r_{14} + r_{34} + r_{23} + r_{12} & = & 2 \\
\label{qua5} r_{45} + r_{34} + r_{23} + r_{25} & = & 2 \\
\label{cu1} r_{15} + r_{35} + r_{13} & = & 2 \\
\label{cu2} r_{45} + r_{24} + r_{25} & = & 2 \\
\label{cu3} r_{14} + r_{34} + r_{13} & = & 2 \\
\label{cu4} r_{25} + r_{35} + r_{23} & = & 2 \\
\label{cu5} r_{14} + r_{24} + r_{12} & = & 2 
\end{eqnarray}

At this point we are in a situation where we have ten unknown R-charges and seventeen equations to satisfy, the five beta functions and the twelve R-charge equations. The vanishing of the beta functions imposes a linear dependence on them reducing the total number of equations to sixteen while some of the R-charge conditions are linearly dependent on others. In order for the system to have a solution, one has to choose subsets of R-charge relations of rank six. 

Therefore, for Type I five-block quivers one has to make a choice of subsets of gauge invariant operators to contribute to the superpotential. 
The choice can be made by suitably adjusting the couplings of the rest of the operators to zero. 
In other words, in the five block case superconformality imposes some form of hierarchy among the couplings of the theory.
Mathematically, this is reflected by the fact that not all of the equations \eqref{relations1} are linearly independent and they cannot all be satisfied simultaneously. 

In total, there are 33 choices of subsets of rank six, each of which can be solved consistently.
We list these sets in Appendix \ref{subsets1}. Note that each subset is required to have cardinality at least, but not exactly, six since some of the relations may be linearly dependent on others. 
For example the set of R-charge relations number $(33)$ of the collection \eqref{subsets1} picks out the relations \eqref{qui1},\eqref{qua1},\eqref{qua2},\eqref{qua3},\eqref{qua4},\eqref{qua5}. 
The cubic relations are linearly dependent on these so for this choice one has to set to zero only the coupling of the quintic operator\footnote{Note that a term like \eqref{yqui11} represents a collection of operators in the superpotential since there are more than one nodes in each block.} \eqref{qui2}.
\begin{equation}\label{yqui11}
y_{13524}\hat{a}_{13} \hat{a}_{35} \hat{a}_{52} \hat{a}_{24} \hat{a}_{41}  \ .
\end{equation}
That is, 
\begin{equation}
y_{13524}=0.
\label{zero33}
\end{equation}
Doing this enables one to bypass the marginality condition \eqref{qui2} because this quintic term decouples from the system which now admits a solution. 
Putting together the requirement of the vanishing of the weighted sum of the beta functions $\sum N_i\beta_i=0$, the anomaly cancellation \eqref{kernel1} and the chosen set of marginal operators, we obtain a Diophantine equation in terms of the quiver data. 
We will take advantage of the discussions above and cast the equation into a quadratic form:
\begin{equation}\label{Diophantos1}
\frac{A_1^2}{\alpha_1} + \frac{A_2^2}{\alpha_2} + \frac{A_3^2}{\alpha_3}  + \frac{A_4^2}{\alpha_4} + \frac{A_5^2}{\alpha_5} = a_{12}A_1A_2 + a_{34}A_3A_4 + a_{51}A_1A_5 + a_{52}A_2A_5 \ ,
\end{equation}
where we recall $A_i$ from \eqref{kernel1} 
and, in fact, $A_iA_j=C_{ij}\equiv(-)^{i+j}M_{ij},$ where $C_{ij}$ is a cofactor and $M_{ij}$ is the $\{i,j\}$ minor of the reduced quiver matrix $q_5$. 
The RHS of the above equation can be written in 5 equivalent ways,
\begin{equation}\label{Dio1inv}
\begin{split}
 & a_{34}A_3A_4 + a_{51}A_1A_5 + a_{12}A_1A_2 + a_{52}A_2A_5  \\ & a_{45}A_4A_5 + a_{51}A_1A_5 + a_{23}A_2A_3 + a_{41}A_1A_4  \\ &
a_{34}A_3A_4 + a_{45}A_4A_5 + a_{12}A_1A_2 + a_{35}A_3A_5  \\ & a_{34}A_3A_4 + a_{51}A_1A_5 + a_{23}A_2A_3 + a_{24}A_2A_4  \\ &
a_{45}A_4A_5 + a_{12}A_1A_2 + a_{23}A_2A_3 + a_{13}A_1A_3 \ .
\end{split}
\end{equation}
Re-organizing, as before in the three-block case, we can rewrite \eqref{Diophantos1} as a sum over minors $M_{ij}$ of $q_5$:
\begin{equation} \label{Dio1co}
\sum_{i} C_{ii} \prod_{j \neq i}\alpha_j - \sum_{i<j} q_5(i,j) C_{ij} \prod_{k} \alpha_k = 0 \ ,
\end{equation}
where the indices run in $\{1,2,3,4,5\}$. 
Upon considering the fact that the determinant of $q_5$ (being an antisymmetric matrix of odd dimension) is zero and that the determinant can be expressed as an alternating sum of minors along any line or column weighted by the respective matrix elements, eq. \eqref{Dio1co} reduces to \eqref{Diophantos1} with the RHS being any of \eqref{Dio1inv}.
Written in this way, this five-block equation is a straightforward generalization of the one found for the three-block case. 

However, this formula is correct only for this specific subset of operators and the ones that are related to it by permutations and arrow reversals as we will see in subsection \ref{equiv-type-1}. This subset is special in the sense that it is the one with maximal cardinality. In other words it is the one that requires the lowest number of couplings to be set to zero. As we previously saw, the only such coupling for this choice of simultaneously marginal operators is \eqref{zero33}.

In this case, one can write the Diophantine equation as the ``signed'' Tits form of the quiver in complete analogy with the three-blocks:
\begin{equation}\label{tits5b}
q_{Q_s}(x_1,x_2,x_3,x_4,x_5)=\sum_{i} \alpha_i x_i^2-\sum_{i<j}q_5(i,j)\alpha_i\alpha_j x_ix_j \ .
\end{equation}
In other words the dimension vectors for which the resulting gauge theory is superconformal satisfy 
\begin{equation}\label{tits5breal}
q_Q(x_1,x_2,x_3,x_4,x_5)=-2\sum_{i<j\;|\;q_5(i,j)<0}|q_5(i,j)| \alpha_i \alpha_j x_ix_j \;,
\end{equation}
where $q_Q$ is the actual Tits form \eqref{tits-form} of the representation.
Upon setting 
$x_j\propto\sqrt{\frac{\prod_{i\neq j}\alpha_i}{\alpha_j}}A_j$
one arrives at equation \eqref{Dio1co}. The observation that the equation can be written as a sum of minors now stems from the Tits form construction. The robustness of our results for the low block numbers implies that they hold for any block quiver. Before formalizing this conjecture let us dwell more on this specific case.

Furthermore, one can go on and solve for the R-charges. Solving for the 4 out of 5 beta-functions in addition to the R-charge marginality conditions we have 10 equations and 10 unknowns. The fifth beta-function will vanish by construction since we have also imposed the Diophantine equation. We find the following rational functions:
\begin{eqnarray}\label{rational-charges}
r_{12} &=& \frac{2}{A_1A_2}\Big( a_{45} \frac{A_3}{\alpha_3} - (\alpha_4 a_{34}a_{45} + a_{35})\frac{A_4}{\alpha_4} + a_{34}\frac{A_5}{\alpha_5} \Big) \nonumber\\
r_{13} &=& \frac{2}{A_1A_3}\Big( -a_{45}\frac{A_2}{\alpha_2} + a_{23}a_{45} A_3 + (\alpha_4 a_{24}a_{45} - a_{52})\frac{A_4}{\alpha_4} - a_{24} \frac{A_5}{\alpha_5} \Big) \nonumber\\
r_{14} &=& \frac{2}{A_1A_4}\Big( -a_{35}\frac{A_2}{\alpha_2} + (\alpha_3 a_{23}a_{35} - a_{52})\frac{A_3}{\alpha_3} +a_{23}a_{45} A_4 - a_{23} \frac{A_5}{\alpha_5} \Big) \nonumber\\
r_{15} &=& \frac{2}{A_1A_5}\Big(  a_{34}\frac{A_2}{\alpha_2} -(a_{24} +\alpha_3 a_{23}a_{34} )\frac{A_3}{\alpha_3} + a_{23}\frac{A_4}{\alpha_4}  \Big) \nonumber\\
r_{23} &=& \frac{2}{A_2A_3}\Big( a_{45}\frac{A_1}{\alpha_1} + a_{51}\frac{A_4}{\alpha_4} - (\alpha_5 a_{45}a_{51} + a_{41})\frac{A_5}{\alpha_5} \Big) \\
r_{24} &=& \frac{2}{A_2A_4}\Big( -a_{35}\frac{A_1}{\alpha_1} + a_{51} \frac{\alpha_3 a_{34}A_4 +\alpha_3 a_{35}A_5 - A_3}{\alpha_3}  - a_{13}\frac{A_5}{\alpha_5} \Big) \nonumber\\
r_{35} &=& \frac{2}{A_3A_5}\Big( -a_{24}\frac{A_1}{\alpha_1} - a_{41}\frac{A_2}{\alpha_2} + a_{12} \frac{\alpha_4 a_{24}A_2 +\alpha_4 a_{34}A_3 - A_4}{\alpha_4}  \Big) \nonumber\\
r_{45} &=& \frac{2}{A_4A_5}\Big( a_{23}\frac{A_1}{\alpha_1}  - (\alpha_2 a_{23}a_{12} + a_{13})\frac{A_2}{\alpha_2} + a_{12} \frac{A_3}{\alpha_3} \Big), \nonumber
\end{eqnarray}
with the remaining 2 given by the marginality conditions \eqref{qui1} and \eqref{qua1}-\eqref{qua5}. The rationality of these values is not surprising since we solved a linear system of ten equations in ten variables. A comment though is in order at this point. There are known five-block del Pezzo quivers for which a-maximization predicts irrational R-charges in contrast with \eqref{rational-charges}. The subtlety lies in the fact that for these models the R-charges of fields between two blocks are not the same and the formulas  \eqref{rational-charges} cannot be applied. Nevertheless, as we will see in the next section, these cases are still solutions of a reduced version of \eqref{Dio1co}.
\paragraph{Comments on multitrace operators}

Let us here briefly comment on the possibility of multitrace operators of the form $(\hat{X}_{12}\hat{X}_{23}\hat{X}_{31})^m,$ for the three block models of sec.~\ref{sec:3b}, and analogously for the five block ones. Such terms, although generically irrelevant, may acquire large anomalous dimensions due to strong coupling effects and become marginal. The R-charge condition \eqref{3r} would then become
\begin{equation}\label{3rn}
r_{12}+r_{23}+r_{31} = \frac{2}{m},
\end{equation} 
and the same change would apply to the R-charge conditions for five block models \eqref{relations1}.
This leads to the following three block Diophantine equation 
\begin{equation} \label{markovn}
\frac{a_{23}^2}{\alpha_1}+\frac{a_{31}^2}{\alpha_2}+\frac{a_{12}^2}{\alpha_3} = \frac{2m-1}{m} a_{12}a_{23}a_{31} \ .
\end{equation}
This equation can be cast into a minor formula as
\begin{equation} \label{comarkoffn}
m \sum_{i} C_{ii} \prod_{k \neq i}\alpha_k  - (1-2m) \sum_{i<j} q_3(i,j) C_{ij} \prod_k \alpha_k = 0 \ ,
\end{equation}
which correctly reduces to \eqref{comarkoff} for $n=1$. For the five block case, the minor formula can be written as
\begin{equation} \label{Dio1con}
m \sum_{i} C_{ii} \prod_{j \neq i}\alpha_j - \sum_{i<j} \left(\frac{1+(-)^{i+j+1}}{2}(1-2m) + \frac{1+(-)^{i+j}}{2}(2-3m) \right) q_5(i,j) C_{ij} \prod_{k} \alpha_k = 0 .
\end{equation}
For $m=m_*=1$ this formula reduces to \eqref{Dio1co}, and in that case, since $1-2m_* = 2-3m_*$, this is a straightforward generalisation of the three block one \eqref{comarkoffn}, but for higher values this is not true. Furthermore, although these equations can be written as a deformed Tits form, the representation theoretic meaning of such an object would be unclear. Interestingly, \eqref{comarkoffn} for the three block models leads to a whole new family of solutions that differ from the ones found in \cite{Benvenuti:2004dw} for the case $m=1$, however, a systematic approach to these cases evades the scope of this paper.

\subsubsection{Reproducing Known Theories}
Now, since we are doing a classification of consistent block quivers which might admit superconformal fixed points, we need to check whether theories known in the AdS/CFT literature are special cases.
In this subsection we verify that the toric quiver gauge theories constructed in \cite{Hanany:2012hi,Davey:2009bp} are indeed a subclass of solutions of the Diophantine equation presented here. 
These models, being toric, have the same rank $N$ in all blocks. 

\begin{figure}[tbh!!]
\begin{center}
\includegraphics[scale=0.25]{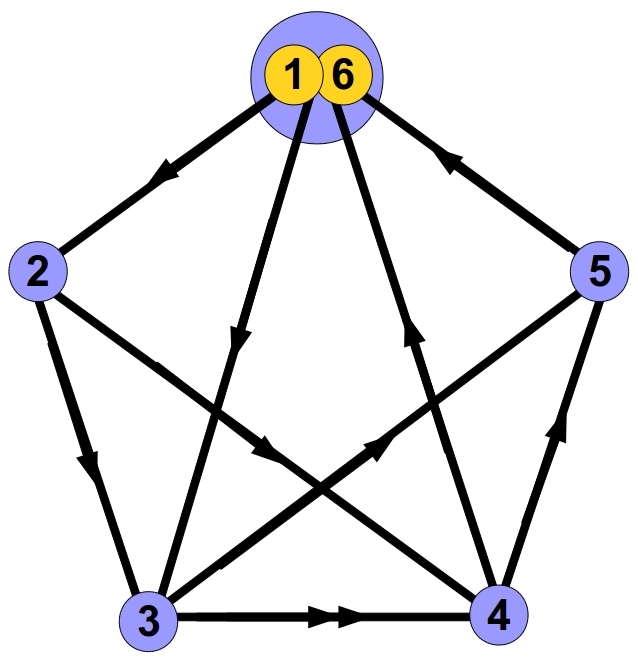}   
\end{center}
\caption{The quiver of the second toric phase of the del Pezzo 3 theory.}
\label{fig:dp3}
\end{figure}

The requirement of equal ranks translates into setting $x=\frac{A_1}{\alpha_1}=\frac{A_2}{\alpha_2}=\frac{A_3}{\alpha_3}=\frac{A_4}{\alpha_4}=\frac{A_5}{\alpha_5}$. Then the rank of the blocks decouples from the equations as a free parameter and the anomaly cancellation condition \eqref{kernel1} becomes
\[
q_5\cdot (\alpha_1,\alpha_2,\alpha_3,\alpha_4,\alpha_5)^\top ={\mathbf 0} \ . 
\]
Since the block multiplicities should lie in the kernel of the adjacency matrix, we must replace $A_i$ with $\alpha_i$. Then equation \eqref{Dio1co} reads 
\begin{equation}\label{Dio1toric}
\sum_{i=1}^5 \alpha_i - \alpha_1 \alpha_2 a_{12} - \alpha_1 \alpha_3 a_{13} + \alpha_1 \alpha_4 a_{41} - \alpha_2 \alpha_3 a_{23}  - \alpha_2 \alpha_4 a_{24} - \alpha_3 \alpha_4 a_{34}  = 0 \ ,
\end{equation}
with the rest of the arrows given by the relations
\begin{eqnarray}\label{dio1toricrels}
a_{51} & = & \frac{\alpha_2 a_{12} + \alpha_3 a_{13} - \alpha_4 a_{41} }{\alpha_5} \quad , \quad a_{52} = \frac{\alpha_3 a_{23} + \alpha_4 a_{24} - \alpha_1 a_{12} }{\alpha_5}\nonumber\\
a_{35} & = & \frac{\alpha_1 a_{13} + \alpha_2 a_{23} - \alpha_4 a_{34}}{\alpha_5} \quad , \quad a_{45} = \frac{\alpha_2 a_{24} + \alpha_3 a_{34} - \alpha_1 a_{41}}{\alpha_5}.
\end{eqnarray}
As can be seen from the quiver diagram these relations are nothing but the requirement of having equal number of incoming and outgoing arrows for each block. The conditions \eqref{dio1toricrels} where chosen randomly on block five since they fix all its incident arrows in terms of the others. Given these substitutions, all the cofactors of the new adjacency matrix equal the cofactor $C_{55}$ of the initial one. This cofactor is the determinant of the four-block matrix that is obtained by deleting the fifth row and column of $q_5$, or in other words it represents the four-block model that arises from the five-block one when we remove the fifth node. For all the known five block models in the literature, this determinant vanishes so that the four block sub-quivers are anomaly free. The equality of all the cofactors ensures that this will then be valid for all the $4\times 4$ sub-determinants representing all the $4b$-models that can arise by the removal of a node. 
By imposing the anomaly cancellation for the sub-quivers, we essentially further reduce the rank of the matrix from $r[q_5]=4$ to $r[q_5]=2$, since we impose relations for every sub-determinant to vanish. The kernel space of such a matrix is therefore 3 dimensional and an arbitrary vector reads
\begin{equation}\label{ker-reduced}
\left(\begin{array}{c} \alpha_1 \\ \alpha_2 \\ \alpha_3 \\ \alpha_4 \\ \alpha_5 \end{array}\right) = \alpha \left(\begin{array}{c}-a_{52} \\ a_{51} \\ 0 \\ 0 \\ a_{12}\end{array}\right) + \beta\left(\begin{array}{c}a_{24} \\ a_{41} \\ 0 \\ a_{12} \\ 0\end{array}\right) + \gamma\left(\begin{array}{c}a_{23} \\ -a_{13} \\ a_{12} \\ 0 \\ 0 \end{array}\right),
\end{equation}
with $\alpha,\beta,\gamma$ positive integers.
Given these substitutions for the block multiplicities the Diophantine equation finally reads
\begin{eqnarray}\label{Dio1toric-reduced}
&&\alpha \Big[-a_{52}(\alpha a_{12}a_{51}-1) + a_{51}(\beta a_{12}a_{24}-1) + a_{12}(\gamma a_{51}a_{23}-1)  \Big]+ \nonumber\\ && 
 \beta \Big[a_{12}(\alpha a_{51}a_{24}-1) + a_{24}(\beta a_{12}a_{41}-1) +a_{41}(\gamma a_{12}a_{23}-1)\Big]+
\\ && 
 \gamma \Big[a_{23}(\alpha a_{51}a_{12}-1) + a_{12}(\beta a_{23}a_{41}-1) -a_{13}(\gamma a_{12}a_{23}-1)\Big] = 0, \nonumber
\end{eqnarray}
with the rest of the arrows given by
\begin{equation}\label{dio1toricrels-reduced}
a_{34} = \frac{ a_{24}a_{13} + a_{23}a_{41} }{a_{12}}, \quad a_{35} = \frac{a_{23} a_{51} - a_{13}a_{52} }{a_{12}},\quad
a_{45} = \frac{a_{24}a_{51} + a_{41}a_{52}}{a_{12}}.
\end{equation}
The known del Pezzo quivers $PdP_2$ , $dP_2^{II}$ , $dP_3^{II}$ , $PdP_{3b}^{II}$ , $PdP_4$ are solutions of equations \eqref{ker-reduced},\eqref{Dio1toric-reduced},\eqref{dio1toricrels-reduced} with $(\alpha,\beta,\gamma)=(1,1,1)$. For example denoting the solution vector of a model as $$(\alpha_1,\alpha_2,\alpha_3,\alpha_4,\alpha_5;a_{12},a_{13},a_{41},a_{51},a_{23},a_{24},a_{52},a_{34},a_{35},a_{45})$$the quiver of the second toric phase of the $dP_3$ theory (see Fig.~\ref{fig:dp3}) corresponds to $${\mathbf v}_{dP_3}^{II}=(2,1,1,1,1;1,1,1,1,1,1,0,2,1,1).$$
For the sake of completeness we present the other solutions of the known superconformal del Pezzo quivers.
\begin{eqnarray*}
{\mathbf v}_{PdP_2} &=& (1,1,1,1,1;2,0,1,1,2,1,1,1,1,1) \\
{\mathbf v}_{dP_2}^{II} &=& (1,1,1,1,1;1,1,1,1,2,0,1,2,1,1) \\
{\mathbf v}_{PdP_{3b}}^{II} &=& (2,1,1,1,1;1,1,1,1,1,1,0,2,1,1)\\
{\mathbf v}_{PdP_4} &=& (1,1,2,2,1;1,1,1,1,1,0,1,1,0,1)
\end{eqnarray*}

\subsubsection{Equivalence Classes for Type I}\label{equiv-type-1}
As previously mentioned, the five-block case is the first where one has to make a choice of simultaneously marginal operators in the superpotential. For Type I there are 33 such subsets of six linearly independent R-charge relations, listed in the Appendix \eqref{subsets1}, which lead to 33 Diophantine equations.
Are any of these equivalent to each other?
The answer is positive but unfortunately not for all. Before discussing that let us clarify what is the equivalence relation.
 
A $n$-ary quadratic form can be represented by a symmetric $n \times n$ matrix as
\begin{equation}
q({\mathbf x})=\sum_{i,j=1}^n a_{ij}x_ix_j\equiv \frac{1}{2}{\mathbf x}^T C_q {\mathbf x}
\end{equation}
where ${\mathbf x}$ is a column vector with entries $x_1\ldots x_n$ and $C_q$ is the symmetric matrix with elements $$C_q(i,j)=(a_{ij}+a_{ji})$$ Two forms are equivalent if their corresponding symmetric matrices are related by a similarity transformation. This is because if $C_q=S^T \cdot C_{q'} \cdot S$ then all the values of $q'$ are determined from the values of $q$ as $q({\mathbf x}) =q'(S \cdot {\mathbf x})$. All the Diophantine equations can be written as quadratic forms over the variables $A_1,\ldots,A_5$ defined in \eqref{kernel1}. We thus consider as equivalent two Diophantine equations, corresponding to two different choices of R-charge conditions, if they are equivalent as quadratic forms. This means that if one solution ${\mathbf x_0}$ of the former equation is known then a solution of the latter can be immediately written as ${\mathbf y_0} = S \cdot {\mathbf x_0}.$

We find that there are 6 equivalence classes. Although the Diophantine equations are written as quadratic forms, the do not obey the nice structure of summation over minors, neither are their Tits forms negative definite. Since this analysis does not give any further insight into what we have already seen, we list all our results in Appendix \ref{app:equiv-I}. In there the reader can find the Diophantine equations for each representative choice of R-charge relations for the Type I quivers. For each choice we also identify which couplings of the superpotential must be set to zero.

\subsubsection{Enumeration of Other Types}
We recall from the beginning of the section that there are 6 distinct, topologically inequivalent, types of five-block quivers and we discussed Type I in detail above.
Fortunately, all the other five Types of quiver diagrams and their equivalence classes, are related to Type I by permutations and orientation reversal operations on the arrows. In other words, all the Diophantine equations obtained from the various subsets of R-charge marginality conditions for each Type are equivalent to those of Type I. 

We find that Type II, III and V lead to a unique set of simultaneously marginal operators and are equivalent with Class 4 of Type I, represented by the set (33) of R-charge conditions, which we discussed in detail in the previous subsections. Types IV and VI have 22 and 11 sets of simultaneously marginal operators which lead to the same numbers of equations, again related to the various classes of Type I. The cycle structure of these quivers are listed in Appendix \ref{other-types}.  

\subsubsection{Duality Tree for Five-Block Models}
Let us now comment on Seiberg duality and see that indeed it leaves our Diophantine equation invariant. As we saw in section \ref{t1} the five-block Diophantine equation is a straightforward generalization of the three-block one, and thus Seiberg duality is easily seen to correspond to affine Weyl reflections for the five-block models as well. 
Let us here focus on the duality as the set of operations reported in \cite{Feng:2001bn} and reviewed in Section \ref{sec:3b}. As a bi-product we will find out that the duality exchanges equations among the six quiver Types that we have.  

\begin{figure}[tbh!!!]
\centering
\includegraphics[scale=0.45]{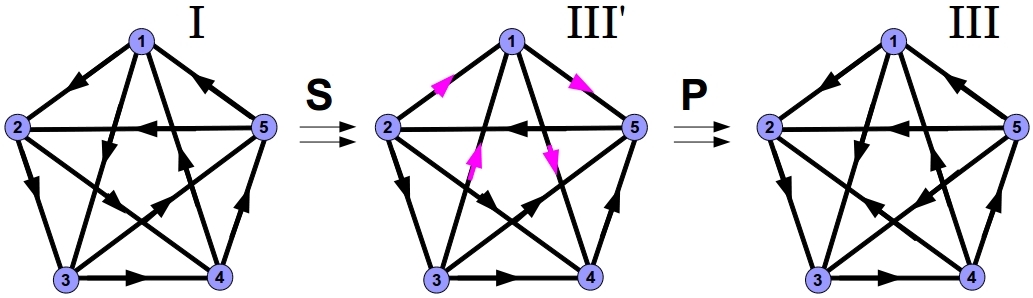}
\caption{{\sf Seiberg duality acting on a quiver of type I. Green arrows are the ones that change direction. $\mathbf{P}$ corresponds to the permutation that brings the middle quiver to its canonical form as it is defined in Fig.~\ref{fig:6types}. The top right labels denote the type of each diagram.}}
\label{fig:seiberg5b}
\end{figure}

So let us analyze a specific example for the Type I quiver drawn in Fig.~\ref{fig:type1} where, for simplicity,  all the block multiplicities are set to one. Suppose we want to dualize with respect to node ``1" in the quiver. This amounts to the following operations
\begin{equation}
\begin{split}
&a_{51}\rightarrow -a_{51},\;a_{41}\rightarrow -a_{41},\;a_{13}\rightarrow -a_{13},\;a_{12}\rightarrow -a_{12}\\
&a_{24}\rightarrow a_{24}-a_{12}a_{41},\;a_{52}\rightarrow a_{52}+a_{12}a_{51},\;a_{34}\rightarrow a_{34}-a_{13}a_{41},\;a_{35}\rightarrow a_{35}-a_{13}a_{51}
\end{split}\label{seibergnode1}
\end{equation}
Let us assume that  
\begin{equation}a_{24}>a_{12}a_{41}, \; a_{52}>a_{12}a_{51}, \; a_{34}>a_{13}a_{41}, \; a_{35}>a_{13}a_{51},
\label{dualarrows}
\end{equation} so that the ``dual" arrows do not change direction. Then Seiberg duality leads to the quiver in the middle quiver of Fig.~\ref{fig:seiberg5b}.
Note that the reversal of the arrows incident to node ``1'' changes the cycle structure of the quiver. In order to see of what Type is the dualized graph, one has to count its oriented cycles. By doing so we find that it is of Type III. The permutation $P=(1)(24)(35)$ brings the dualized diagram to its canonical form as defined in Fig.~\ref{fig:6types}, that is with a clear counter-clockwise orientation of the ``outer" pentagon (the perimeter). Note that had we violated one of \eqref{dualarrows} we would have an outcome of another type.
Recall that from the analysis in section \ref{t1} the rank of the node $``i"$ is proportional to $A_i$ as in \eqref{kernel1}. The transformations \eqref{seibergnode1} act on the $A$'s as follows
\begin{equation}
A_1\rightarrow A_1 - a_{41} A_4 - a_{51} A_5,\; A_{2}\rightarrow -A_2,\;A_{3}\rightarrow -A_3,\;A_{4}\rightarrow -A_4,\;A_{5}\rightarrow -A_5.
\label{dualranks}
\end{equation}
The minus signs in front of the dual ranks are notational artifacts since they arise due to the fact that we consider the arrows to change sign when reversed. The anomaly cancellation forces the dimension vector to be in the kernel of the quiver matrix. The overall minus in the dual ranks is due to the fact that the duality as we define it on the quiver data reflects this vector through the origin of the null space of $q_5$. This is also evident in the three-block case where the rank vector is proportional to the vector with entries the non incident arrow of each node (cf. \eqref{ker3}). Had we ended up with a minus sign in the rank of some blocks and positive in the others, then we would face a real problem, which is certainly not the case here.

It is straightforward to verify that the transformation $A_1\rightarrow \alpha_1 a_{41} A_4 + \alpha_1 a_{51} A_5 -  A_1$ leaves the five-block Diophantine equation invariant! The best way to see this is to use the last line of \eqref{Dio1inv} as the RHS of \eqref{Diophantos1} since it involves only arrows that remain unaffected from the operations \eqref{seibergnode1}. This corresponds exactly to $N_{C_1} \mapsto N_{F_1} - N_{C_1}$. The fact that the determinant vanishes ensures that the number of flavors is uniquely defined, that is $ a_{41} A_4 + a_{51} A_5 = a_{12} A_2 + a_{13} A_3$.

\subsection{Summary and Generalization to $n$-Blocks}
In this section we have presented our results for block models up to five nodes, including the known five-block theories on del-Pezzo surfaces. We saw that they are underlined by an identical algebraic structure as the three-block ones, while for four-blocks the situation is slightly altered. Even though there exists a formula which admits a similar structure as in the odd cases the connection with representation theoretic concepts is blurry. Finally, we saw how Seiberg duality can be realized as an action on the $5b$-quiver that leaves the Diophantine equation invariant.


Unfortunately, due to the exponential increase in complexity we cannot explicitly verify  higher-block quivers but the persistence of the summation over minors formula strongly recommends a continuation to any quiver.
Our highly non-trivial analysis and the robustness of our results for the low numbers of blocks leads us to conjecture a generic classification of chiral quiver gauge theories satisfying the necessary conditions for an ${\cal N} =1$ superconformal fixed point to exist:

\begin{conjecture}
{\it Given a quiver with $n=2l+1$ blocks and the maximal set of simultaneously marginal operators, the resulting anomaly free theory has vanishing beta and gamma functions if the quiver data, in the notation of Sec.\ref{notation} (cf. p.6), satisfy the following Diophantine equation:
\begin{equation} \label{odd-blocks}
\sum_{i} C_{ii} \prod_{j \neq i}\alpha_j - \sum_{i<j} q_n(i,j) C_{ij} \prod_{k} \alpha_k = 0 \ ,
\end{equation}
where $C_{ij}$ is the $(i,j)$ cofactor of the anti-symmetrized adjacency matrix $q_n$ and the indices run in $\{1,\ldots,n\}$. \\
The dimension vector corresponding to the gauge theory satisfies
\begin{equation}\label{conj}
q_Q(x_1,...,x_n)=-2\sum_{i<j \;|\; q_n(i,j)<0}|q_n(i,j)| \alpha_i\alpha_j x_ix_j ,
\end{equation}
where $q_Q$ is the Tits form of the quiver.
The rank of block $i$ is given by
\begin{equation}\label{tits-to-markovnb-conj}
x_i\propto\sqrt{\frac{M_{ii} \prod_{j\neq i}\alpha_j}{\alpha_i}},
\end{equation}
with $M_{ij}$ being the $(i,j)$, $(n-1)\times (n-1)$ first minor of $q_n$. The proportionality constant is fixed so that $x_i\;\in\;\mathbb{Z}$. Therefore, superconformal gauge theories correspond to imaginary roots of the quiver's root system. The affine Weyl group that permutes these roots offers a realization of Seiberg duality in this context.}
\end{conjecture}

\begin{conjecture}
{\it Given a quiver with $n=2l$ blocks and the maximal set of simultaneously marginal operators, the resulting anomaly free theory has vanishing beta and gamma functions  if the quiver data, satisfy the following Diophantine equation:
\begin{eqnarray}\label{even-form}
\sum_{i_{_1}<i_{_2}} M_{i_{_1}i_{_2},i_{_1}i_{_2}} \prod_{m\neq i_{_1},i_{_2}}\alpha_m &-& \sum_{i_1\neq i_2<i_3} (-)^{i_2+i_3} q_n(i_2,i_3) M_{i_1i_2,i_1i_3} \prod_{m\neq i{_1}}\alpha_m + \nonumber \\ &&\qquad\qquad + \sum_{i_{_1}<\ldots <i_{_n}} q_n(i_{_1},i_{_2})q_n(i_{_1},i_{_n}) M_{i_{_1}i_{_2},i_{_1}i_{_n}} \prod_{m}\alpha_m = 0
\end{eqnarray}
where $M_{ij,kl}$ is the $(ij;kl)$ second minor of the anti-symmetrized adjacency matrix $q_n$ and the indices run in $\{1,\ldots,n\}$. \\
Having a superconformal $2l$-block model the operation of removing one block leads to a superconformal $(2l-1)$-block model; this does not hold for $(2l+1)$-block quivers reduced to $2l$-block theories.

}
\end{conjecture}
Even though we have calculated explicitly one even-block quiver the way that $4b$ reduces to $3b$ through the minor formula is suggestive for the $6b$ construction as well. Note that although the last term in \eqref{even-form} does not participate in the $2l-1$ quiver, since it is weighted by all the node multiplicities $\alpha_m$, including the one we set to zero in order to descent to $2l-1$-blocks, the way it is written it reproduces a $2l$-order term in the arrow multiplicities, which is the order $2l$ operator in the perimeter of the polygon. That is for example the analogue of the term $a_{12} a_{23} a_{34} a_{14}$ in \eqref{4b-ami}. This is because the second minor of a $2l\times 2l$ matrix is of order $2l-2$ in the entries and together with the quadratic piece yields the term of order $2l$. 


\section{Conclusions and Outlook}\label{s:conc}
In this paper we have organized chiral quiver theories into block structure and derived the necessary conditions for a theory to be superconformal.
The pigeon-holing of the plethora of quiver theories into block-quivers
each block of which contains non-adjacent nodes dramatically reduces the complexity of the problem and gives a handle on
a step-wise catalogue of the gauge theories of this class that might admit a superconformal fixed point. Many of the complicated geometries which ordinarily give
rise to product gauge groups, especially Calabi-Yau
manifolds as cones over higher del Pezzo surfaces, now simply belong to the class of 3-, 4- or 5-block models.
Importantly, we have incorporated physical conditions of anomaly-cancellation, conformality as well
as marginality by enforcing all superpotential terms to have R-charge 2, directly into our scheme;
these strong constraints translate to combinatorics.
We envision that with further computer work, we can efficiently classify more and more of quivers with superpotential.

A powerful invariant for a block-quiver is an underlying Diophantine equation which the adjacency matrix and the ranks of the
nodes must obey. Interpreting the ranks as the dimension vector of the representation of the quiver, and using the so-call
Tits quadratic form thereon, we have shown how the Diophantine equation arises upto 5-blocks
and conjectured a general form. 
The exponential increase in complexity of our problem as the number of blocks rises, forbids us to further support our results by direct computation, but we find it highly non trivial and suggestive that the first three cases can be attacked in a unified way. The complication of having to enumerate the inequivalent graphs for higher block number is one more computational obstacle. The first few terms in the sequence that counts inequivalent graphs up to seven blocks are $(0,0,1,1,6,36,356,...)$.

For each geometry, there is a tree of Seiberg-dual theories arising by consecutive action on the various nodes (blocks). The ranks and 
subsequent adjacency matrices of these dual theories are, surprisingly, controlled by this Diophantine equation by precisely being
its solutions. Therefore understanding of this equation is of great significance.

Taking advantage of the presentation of the Diophantine equation as the Tits form of the quiver, Seiberg duality is seen as affine Weyl reflections in the space of roots, provides a representation-theoretic approach -
complementing the usual geometric ones such as mutation and Picard-Lefshetz monodromy - to the tree of dualities.
Indeed, the Diophatine equation is invariant under such Weyl reflections. Furthermore, this point of view draws a connection among the representation theory of quivers, their root systems and $\mathcal{N}=1$, 4-d superconformal gauge theories for cases with odd block number. For the quivers with an even number of blocks the situation is more blur. We could not identify a clear connection with a bilinear form but we managed to show that in that case as well, the polynomial invariant that controls possible superconformal fixed points, can be presented as a sum over minors of the adjacency matrix. This fact in combination with the reduction of the even to odd block Diophantine equations suggests that there might be a representation theoretic description of these theories as well, a topic we leave for future work.

New mathematical descriptions of physical phenomena is beneficial for both fields. For example, realizing field theory dualities in different mathematical contexts may reveal aspects thereof not previously accessible and even lead to new connections among field theories. Alternatively, connecting root systems of wild quivers whose representation theory is unknown, to physical systems such as conformal field theories may open a new path of studying such objects. Our construction, even though restricted to a subclass of quiver theories and their superconformal fixed points, provides such a link by mapping field theories obeying a set of physical requirements to subsets of imaginary roots of the quiver, and thus takes a step towards this direction.



In addition to fitting further known theories into our context, which also includes a huge class of known examples, there is much
left to do.
Given the conjectural forms of the Diophatine equations and block structures, we can reverse engineer the subsequent quivers
with superpotential. We can do this by finding the explicit moduli space of vacua from the quiver data and then find Calabi-Yau
geometries for which the world-volume theories of the D-branes probes are not yet known.
Within the toric sub-class, which comprises most of the known examples to AdS/CFT, there is an interpretation in terms of
brane-tilings where the ranks of all nodes are equal, our classification frame-work thus also gets simplified.
It would be interesting to investigate this class in further detail as well as to further explore, understand and develop our representation theoretic approach to the problem of enumerating superconformal theories and their relation to the underlying root system of the quiver diagram.

\section*{Acknowledgments}
S.S. would like to thank Ibrahim Assem, Sergey Mozgovoy, Alastair King, Konstanze Rietsch and Balazs Szendroi for clarifications and email correspondence and Alexandra Tzirkoti for carefully proofreading the manuscript. He would also like to thank Dimitri Frantzeskakis, Vassos Achilleos and Lia Katsimiga for warm hospitality in University of Athens, Greece where parts of this work were completed. Y.-H.~H. would like to thank the
Science and Technology Facilities Council, UK, for an Advanced 
Fellowship and grant
ST/J00037X/1, the Chinese Ministry of Education, for a Chang-Jiang Chair 
Professorship
at NanKai University, the US National Science Foundation for grant 
CCF-1048082,
as well as City University, London and Merton College, Oxford, for their 
enduring support.

\appendix

\section{Quivers, an Algebraic Interlude} \label{ap:quiver}\setall
In this appendix, we give, in an as self-contained fashion as possible, some rudiments on the representation theory of quiver.
The interested reader is referred to \cite{Crawley,Brion,Savage,Derksen} for a more in depth presentation of this material and to \cite{He:1999xj} for considerations in the gauge-theoretic context.

A quiver diagram is defined as a pair \qui=\pair where \ver is a finite set of vertices and \edg is a finite set of oriented edges connecting these vertices. 
For $\rho\in\mathbf{Q}_1$ we let $h(\rho)$ to denote the vertex attached to the head of the arrow and $t(\rho)$ the one to the tail. 
A path in \qui is a sequence $x=\rho_1\ldots\rho_n$ of arrows such that $h(\rho_{i+1})=t(\rho_i)$. Moreover, for each vertex $i\in\mathbf{Q}_0$ we consider a trivial path $e_i$ which starts and ends in $i$. The {\it path algebra} $kQ$ associated with the quiver is the $k$-algebra whose basis is the collection of paths and with the product rule given by concatenation of the paths and $k$ is some ground number field, usually taken to be $\mathbb{C}$. 
That is, the multiplication is
\begin{equation}
x\cdot y\equiv\Bigg\{\begin{array}{cc} xy,& \text{if}\quad h(y)=t(x)\\
0,&\text{otherwise} \ .
\end{array}
\end{equation}
An important class of quivers consists of the ones that are endowed with a {\it superpotential}. The superpotential is the set of all cyclic paths in the quiver diagram. One can formally define a derivative with respect to arrows, acting on these cyclic paths. The set of derivatives of all cyclic paths with respect to all their constituent arrows forms an ideal called the {\it Jacobian ideal}. The quotient of the path algebra by the Jacobian ideal is referred to as the {\it Jacobian algebra}. We call such a quiver with superpotential a {\it bounded} quiver since it is bounded by zero-relations, while in the absence of a superpotential we refer to the quiver as {\it unbounded}.

Let us illustrate these definitions with two simple examples.

\paragraph*{The Jordan quiver.} 
\begin{figure}[htb]
\includegraphics[scale=0.3]{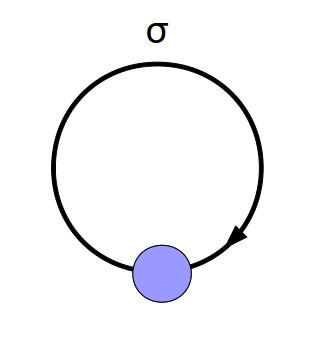}   
\caption{The Jordan quiver.}
\end{figure}
The path algebra of the Jordan quiver is infinite dimensional, with the basis set being $\{e_1,\rho,\rho^2,\rho^3,\ldots\}$. 
The algebra is isomorphic to the polynomial ring $k[t]$.

\begin{figure}
\includegraphics[scale=0.25]{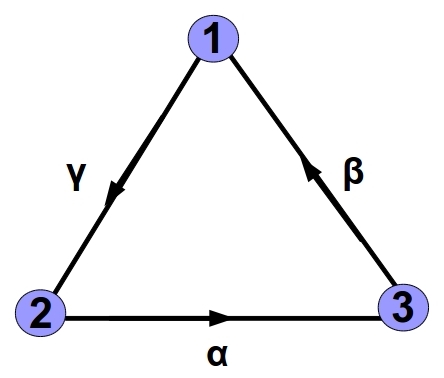}
\caption{The oriented $\hat{A}_2$ quiver.}
\label{fig:3bap}
\end{figure}

\paragraph*{A quiver with relations.}
The path algebra of the quiver depicted in Fig.~\ref{fig:3bap}
has a basis given by the paths $\{e_1,e_2,e_3,\alpha,\beta,\gamma,\beta\alpha,\gamma\beta,\alpha\gamma,\gamma\beta\alpha,\ldots\}$. 
Note that other combinations of arrows are not allowed, for example $\gamma\alpha=0$ since $h(\alpha)\neq t(\gamma)$. This quiver has also a superpotential given by the unique cycle $S=\gamma\beta\alpha.$ The Jacobian ideal is the one generated by the following relations, which form the zero paths,
$$\partial_\alpha S=\gamma\beta \; , \quad \partial_\beta S = \alpha\gamma \; , \quad \partial_\gamma S = \beta\alpha$$

\subsection{Quiver Representations} \label{ap:quiverrjpg}
A representation of a quiver is the assignment of a vector space $V_i$ to each vertex $i\in\mathbf{Q}_0$ and a linear map $V_\rho:\;V_{t_{(\rho)}}\mapsto V_{h_{(\rho)}}$ to each edge $\rho\in\mathbf{Q}_1$.
Different representations of a given quiver are different sets of vector spaces and morphisms that one can assign to each vertex or edge respectively. 
The dimension vector is defined as follows, 
\begin{equation}
d_V=(\dim V_1,\ldots,\dim V_n)\in\mathbb{Z}^{\mathbf{Q}_0}
\end{equation}
where $n$ is the number of vector spaces. This is just the vector labelling the ranks.

Clearly, there are infinite representations, since there are infinite dimension vectors, but one does not need to classify them. 
A key notion is that of indecomposable representations of a given quiver. 
Let $V\equiv(V_i,V_\rho),\;W\equiv(W_i,W_\rho)$ be two representations of a quiver \qui, where Latin indices denote vertices and Greek, edges. 
Define a direct sum of two representations as 
\begin{equation*}
V\oplus W\equiv \Big\{(V\oplus W)_i,(V\oplus W)_\rho\Big\}
\end{equation*}
where the resulting vector space set is
\begin{equation}
(V\oplus W)_i=V_i\oplus W_i
\end{equation}
and the resulting map set $(V\oplus W)_\rho :(V\oplus W)_{t{(\rho)}}\mapsto (V\oplus W)_{h{(\rho)}}$
\begin{equation}
(V\oplus W)_\rho\big((v,w)\big)=\Big(V_{\rho}(u),W_{\rho}(w)\Big),\quad v\in V_{t{(\rho)}},w\in W_{t{(\rho)}}.
\end{equation}
A representation $V$ is \textit{trivial} if $V_i=0,\;\forall\; i\in\mathbf{Q}_0$ and \textit{simple} if its only subrepresentation is the trivial and itself in complete analogy with the group theoretical definitions. In addition, a representation $V$ is \textit{decomposable} if it is isomorphic to $W\oplus U$ for some $W,U \in{\rm Rep}_k(\mathbf{Q})$, and \textit{indecomposable} otherwise.
It is an important fact that
{\it every representation of a quiver diagram has a unique, up to isomorphism, decomposition into indecomposable representations.} 

Thus, one needs only classify the indecomposable representations of a quiver diagram.  

Let us once more illustrate the above notions with two simple examples:
\paragraph*{An unbounded linear quiver.}
\includegraphics[scale=0.4]{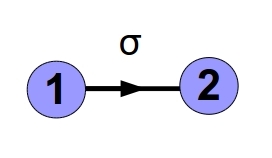}   
This diagram has the following indecomposable representations $U,V,W$, where
\begin{equation}
\{ U_1\cong k,\;U_2=0,\;U_\sigma =0 \}
\ ,
\{ V_1=0,\;V_2\cong k,\;V_\sigma =0 \}
\ ,
\{W_1\cong k,\;W_2\cong k,\;W_\sigma =1 \} \ .
\end{equation}

Therefore, any representation $Z=\{Z_1,Z_2;Z_\sigma\}$ of $\mathbf{Q}$ is isomorphic to \begin{equation}
Z\cong U^\alpha\oplus V^\beta \oplus W^\gamma 
\end{equation} 
with $U^\alpha \equiv\underbrace{U\oplus\ldots\oplus U}_{\alpha}$. 
The positive integers $\alpha,\beta,\gamma$ are related to the rank of the morphism $Z_\sigma$ and the dimensions of the vector spaces $Z_1$ and $Z_2$ as follows.
Since the spaces on the LHS are isomorphic to the direct sum on the RHS for each value of the index $i$, the dimension vectors must be the same. Denoting the dimension of a vector space $A_i$ as ${\rm dim}(A_i)\equiv d_i$, where A runs over all four representations, namely $U,V,Z,W$, we have $$d_1=\alpha+\gamma\quad\text{and}\quad d_2=\beta+\gamma.$$
The exponent $\gamma$ is the rank $\sigma$ of the morphism $Z_\sigma$. Solving the above equations we find $a=d_1-\sigma,\;\beta=d_2-\sigma$. Thus, the decomposition of any representation $Z$ of this quiver, with dimension vector $d_Z=(d_1,d_2)$, is
\begin{equation}
Z\cong U^{d_1-\sigma}\oplus V^{d_2-\sigma} \oplus W^\sigma \ . 
\end{equation} 
Note how $d_Z$ governs the decomposition.
\paragraph*{The bounded $\hat{A}_2$ quiver.}
The quiver with relations depicted in Fig.~\ref{fig:3bap} falls under the category of {\it gentle} algebras \cite{Assem:book,Assem:gentle}. A gentle algebra is defined as the one that has the following properties:
\begin{itemize}
\item[(C1)] At each point of \qui start at most two arrows and stop at most two
arrows.
\item[(C2)] The ideal of zero relations $I$ is generated by paths of length 2.
\item[(C3)] For each arrow $\beta$ there is at most one arrow $\alpha$ and at most one arrow $\gamma$
such that $\alpha\beta \in I$ and $\beta\gamma \in I$.
\item[(C4)] For each arrow $\beta$ there is at most one arrow $\alpha$ and at most one arrow $\gamma$
such that $\alpha\beta \notin I$ and $\beta\gamma \notin I$.
\end{itemize}
The representation theory of these quivers is well studied. Their indecomposable representations fall under two categories, the {\it string modules} and the {\it band modules}. Denoting by $A$ the Jacobian algebra,
a string is by definition a reduced walk $w$ in $A$ avoiding the
zero-relations. A string
is cyclic if the first and the last vertex coincide. A band is defined to be a
cyclic string $b$ such that each power $b^n$
is a string, but $b$ itself is not a proper
power of some string $c$.
The string module $M(w)$ is obtained from the string $w$ by replacing each vertex that belongs to the walk by a copy of the field $k$. The
dimension vector ${\rm dim} M(w)$ of $M(w)$ is obtained by counting how often the
string $w$ passes through each vertex $x$ of the quiver \qui. Similarly, each band $b$ in
$A$ gives rise to a family of band modules $M(b)$. All string and band modules
are indecomposable, and in fact every indecomposable $A-$module is either a
string module $M(w)$ or a band module $M(b)$.
For the $\hat{A}_2$ quiver we have the following string modules: $\{e_1,e_2,e_3\}$ of zero length giving rise to dimension vectors $\{(1,0,0) \; , \; (0,1,0) \; , \; (0,0,1)\}$ and $\{\alpha,\beta,\gamma\}$ of unit length giving rise to dimension vectors $\{(0,1,1) \; , \; (1,0,1) \; , \; (1,1,0)\}$ respectively. Note that there are no band modules since any walk of length greater than one contains the zero relations.

Let us close this section by stating two important theorems on quiver representations:
\par {\bf Gabriel's Theorem.}
 
\begin{itemize}
\item \textit{A quiver is of finite type if and only if the underlying graph is a union of Dynkin graphs of type A,D or E.}
\item \textit {A quiver is of tame type if and only if the underlying graph is a union of Dynkin graphs of type A,D or E  and extended Dynkin diagrams of type $\hat{A},\hat{D}$ or $\hat{E}$.}
\item \textit{The isomorphism classes of indecomposable representations of a quiver \qui of finite type are in one-to-one correspondence with the positive roots of the root system associated to the underlying graph of \qui. The correspondence is given by $$V\mapsto \sum_{i\in\mathbb{Q}_0}d_V(i)\alpha_i$$ where $a_i$ is the $i$-th positive root and by graph is meant the set of edges and vertices without considering the orientations in each case.}
\end{itemize}
\par {\bf Kac's Theorem.}
 \textit{Let \qui be an arbitrary quiver. The dimension vectors of indecomposable representations of \qui correspond to positive roots of the root system of the underlying graph of \qui . Real roots correspond to dimension vectors for which there is exactly one indecomposable representation, while imaginary roots correspond to dimension vectors for which there are families of indecomposable representations. If a
positive root $\alpha$ is real, then $q(\alpha) = 1$. If it is imaginary, then $q(\alpha) \leq 0$.}

\section{Complementary Results}\label{ap:subset6}
In this Appendix, we list the results obtained for the rest of the quivers. We list the cycle structure of each Type as well as all the possible consistent subsets of choices which could satisfy the constraints of the sum of R-charges of each cycle equalling to 2, for Types I, IV and VI of the five-block quiver.
Each set is given as a reference to the equation number in the text and consists of 6 members because, as explained, we need a rank six linear space.
\subsection{Subsets of Marginal Operators for Type I}
There are 33 possible choices:
\begin{eqnarray}\label{subsets1}
{\mathbf{\scriptscriptstyle 1}}\{\eqref{qui2},\eqref{cu1},\eqref{cu2},\eqref{cu3},\eqref{cu4},\eqref{cu5}\} &,& 
{\mathbf{\scriptscriptstyle 2}}\{\eqref{qui2},\eqref{qua1},\eqref{qua3},\eqref{qua4},\eqref{cu1},\eqref{cu2}\} \nonumber\\
{\mathbf{\scriptscriptstyle 3}}\{\eqref{qui2},\eqref{qua1},\eqref{qua3},\eqref{qua4},\eqref{qua5},\eqref{cu1}\}&,& 
{\mathbf{\scriptscriptstyle 4}}\{\eqref{qui2},\eqref{qua1},\eqref{qua3},\eqref{qua5},\eqref{cu1},\eqref{cu5}\}\nonumber\\
{\mathbf{\scriptscriptstyle 5}}\{\eqref{qui2},\eqref{qua1},\eqref{qua4},\eqref{qua5},\eqref{cu1},\eqref{cu2}\}&,& 
{\mathbf{\scriptscriptstyle 6}}\{\eqref{qui2},\eqref{qua1},\eqref{qua2},\eqref{qua3},\eqref{cu2},\eqref{cu3}\}\nonumber\\
{\mathbf{\scriptscriptstyle 7}}\{\eqref{qui2},\eqref{qua1},\eqref{qua2},\eqref{qua3},\eqref{qua4},\eqref{cu2}\}&,& 
{\mathbf{\scriptscriptstyle 8}}\{\eqref{qui2},\eqref{qua1},\eqref{qua2},\eqref{qua3},\eqref{qua4},\eqref{qua5}\}\nonumber\\
{\mathbf{\scriptscriptstyle 9}}\{\eqref{qui2},\eqref{qua1},\eqref{qua2},\eqref{qua3},\eqref{qua5},\eqref{cu3}\}&,& 
{\mathbf{\scriptscriptstyle 10}}\{\eqref{qui2},\eqref{qua1},\eqref{qua2},\eqref{qua4},\eqref{cu4},\eqref{cu5}\}\nonumber\\
{\mathbf{\scriptscriptstyle 11}}\{\eqref{qui2},\eqref{qua1},\eqref{qua2},\eqref{qua4},\eqref{qua5},\eqref{cu2}\}&,& 
{\mathbf{\scriptscriptstyle 12}}\{\eqref{qui2},\eqref{qua1},\eqref{qua2},\eqref{qua5},\eqref{cu2},\eqref{cu3}\}\nonumber\\
{\mathbf{\scriptscriptstyle 13}}\{\eqref{qui2},\eqref{qua3},\eqref{qua4},\eqref{qua5},\eqref{cu3},\eqref{cu4}\}&,& 
{\mathbf{\scriptscriptstyle 14}}\{\eqref{qui2},\eqref{qua2},\eqref{qua3},\eqref{qua4},\eqref{cu1},\eqref{cu2}\}\nonumber\\
{\mathbf{\scriptscriptstyle 15}}\{\eqref{qui2},\eqref{qua2},\eqref{qua3},\eqref{qua4},\eqref{qua5},\eqref{cu1}\}&,& 
{\mathbf{\scriptscriptstyle 16}}\{\eqref{qui2},\eqref{qua2},\eqref{qua3},\eqref{qua5},\eqref{cu1},\eqref{cu3}\}\nonumber\\
{\mathbf{\scriptscriptstyle 17}}\{\eqref{qui2},\eqref{qua2},\eqref{qua4},\eqref{qua5},\eqref{cu1},\eqref{cu2}\}&,& 
{\mathbf{\scriptscriptstyle 18}}\{\eqref{qui1},\eqref{qui2},\eqref{cu1},\eqref{cu2},\eqref{cu4},\eqref{cu5}\}\nonumber\\
{\mathbf{\scriptscriptstyle 19}}\{\eqref{qui1},\eqref{qui2},\eqref{cu1},\eqref{cu2},\eqref{cu3},\eqref{cu4}\}&,& 
{\mathbf{\scriptscriptstyle 20}}\{\eqref{qui1},\eqref{qui2},\eqref{cu1},\eqref{cu2},\eqref{cu3},\eqref{cu5}\}\nonumber\\
{\mathbf{\scriptscriptstyle 21}}\{\eqref{qui1},\eqref{qui2},\eqref{cu1},\eqref{cu3},\eqref{cu4},\eqref{cu5}\}&,& 
{\mathbf{\scriptscriptstyle 22}}\{\eqref{qui1},\eqref{qui2},\eqref{qua1},\eqref{qua3},\eqref{qua4},\eqref{qua5}\}\nonumber\\
{\mathbf{\scriptscriptstyle 23}}\{\eqref{qui1},\eqref{qui2},\eqref{qua1},\eqref{qua5},\eqref{cu3},\eqref{cu5}\}&,& 
{\mathbf{\scriptscriptstyle 24}}\{\eqref{qui1},\eqref{qui2},\eqref{qua1},\eqref{qua2},\eqref{cu2},\eqref{cu5}\}\nonumber\\
{\mathbf{\scriptscriptstyle 25}}\{\eqref{qui1},\eqref{qui2},\eqref{qua1},\eqref{qua2},\eqref{qua3},\eqref{qua4}\}&,& 
{\mathbf{\scriptscriptstyle 26}}\{\eqref{qui1},\eqref{qui2},\eqref{qua1},\eqref{qua2},\eqref{qua3},\eqref{qua5}\}\nonumber\\
{\mathbf{\scriptscriptstyle 27}}\{\eqref{qui1},\eqref{qui2},\eqref{qua1},\eqref{qua2},\eqref{qua4},\eqref{qua5}\}&,& 
{\mathbf{\scriptscriptstyle 28}}\{\eqref{qui1},\eqref{qui2},\eqref{cu2},\eqref{cu3},\eqref{cu4},\eqref{cu5}\}\nonumber\\
{\mathbf{\scriptscriptstyle 29}}\{\eqref{qui1},\eqref{qui2},\eqref{qua3},\eqref{qua4},\eqref{cu1},\eqref{cu4}\}&,& 
{\mathbf{\scriptscriptstyle 30}}\{\eqref{qui1},\eqref{qui2},\eqref{qua3},\eqref{qua5},\eqref{cu1},\eqref{cu3}\}\nonumber\\
{\mathbf{\scriptscriptstyle 31}}\{\eqref{qui1},\eqref{qui2},\eqref{qua2},\eqref{qua3},\eqref{qua4},\eqref{qua5}\}&,& 
{\mathbf{\scriptscriptstyle 32}}\{\eqref{qui1},\eqref{qui2},\eqref{qua2},\eqref{qua4},\eqref{cu2},\eqref{cu4}\}\nonumber\\
{\mathbf{\scriptscriptstyle 33}}\{\eqref{qui1},\eqref{qua1},\eqref{qua2},\eqref{qua3},\eqref{qua4},\eqref{qua5}\}\nonumber
\end{eqnarray}

\subsection{Equivalence Classes for Type I Quivers.}\label{app:equiv-I}
Here we list the the Diophantine equations which represent each class within Type I. Recall that we consider two  equations equivalent if they are equivalent as quadratic forms. The six equivalence classes are
\begin{itemize}
\item[Class 1:] $Dio_1(4),Dio_1(10),Dio_1(12),Dio_1(13),Dio_1(14),Dio_1(22),Dio_1(25),\\
Dio_1(26),Dio_1(27),Dio_1(31)$
\item[Class 2:] $Dio_1(18),Dio_1(19),Dio_1(20),Dio_1(21),Dio_1(28)$
\item[Class 3:] $Dio_1(23),Dio_1(24),Dio_1(29),Dio_1(30),Dio_1(32)$
\item[Class 4:] $Dio_1(1),Dio_1(2),Dio_1(5),Dio_1(6),Dio_1(16),Dio_1(17),Dio_1(33)$
\item[Class 5:] $Dio_1(3),Dio_1(7),Dio_1(9),Dio_1(11),Dio_1(15)$
\item[Class 6:] $Dio_1(8)$
\end{itemize}
where $n$ in $Dio_1(n)$ refers to the $n$-th set of R-charge relations according to the numbering of \eqref{subsets1}. Writing the diagonal part of the quadratic form as $Q_{n}^I=\frac{A_1^2}{\alpha_1} + \frac{A_2^2}{\alpha_2} + \frac{A_3^2}{\alpha_3}  + \frac{A_4^2}{\alpha_4} + \frac{A_5^2}{\alpha_5}$, the representative Diophantine equations are as follows:
\paragraph*{Class 1 : Set 10}

\begin{equation}
Q_{10}^I=\frac{1}{2}{\mathbf A}^T\left(\begin{array}{ccccc}\dfrac{2}{\alpha_1} & a_{12} & 2a_{13} & a_{41} & 0 \\ a_{12} & \dfrac{2}{\alpha_2} & 0 & a_{42} & 0 \\ 2a_{13} & 0 & \dfrac{2}{\alpha_3} & 0 & a_{35} \\ a_{41} & a_{42} &0 & \dfrac{2}{\alpha_4} & 0 \\ 0 & 0 & a_{35} &0 & \dfrac{2}{\alpha_5}  \end{array}\right){\mathbf A}
\end{equation}
\paragraph*{Class 2 : Set 28}

\begin{equation}
Q_{28}^I=\frac{1}{2}{\mathbf A}^T\left(\begin{array}{ccccc}\dfrac{2}{\alpha_1} & 0 & a_{13} & 0 & 3a_{51} \\ 0 & \dfrac{2}{\alpha_2} & 0 & a_{24} & 0 \\ a_{13} & 0 & \dfrac{2}{\alpha_3} & 0 & a_{35} \\ 0 & a_{24} &0 & \dfrac{2}{\alpha_4} & 0 \\ 3a_{51} & 0 & a_{35} &0 & \dfrac{2}{\alpha_5}  \end{array}\right){\mathbf A}
\end{equation}
\paragraph*{Class 3 : Set 32}

\begin{equation}
Q_{32}^I=\frac{1}{2}{\mathbf A}^T\left(\begin{array}{ccccc}\dfrac{2}{\alpha_1} & 2a_{12} & a_{13} & 0 & a_{51} \\ 2a_{12} & \dfrac{2}{\alpha_2} & 0 & a_{24} & 0 \\ a_{13} & 0 & \dfrac{2}{\alpha_3} & 0 & a_{35} \\ 0 & a_{24} &0 & \dfrac{2}{\alpha_4} & 0 \\ a_{51} & 0 & a_{35} &0 & \dfrac{2}{\alpha_5}  \end{array}\right){\mathbf A}
\end{equation}
\paragraph*{Class 4 : Set 1}

\begin{equation}
Q_{1}^I=\frac{1}{2}{\mathbf A}^T\left(\begin{array}{ccccc}\dfrac{2}{\alpha_1} & a_{21} & 0 & a_{41} & 0 \\ a_{21} & \dfrac{2}{\alpha_2} & 0 & a_{24} & 0 \\ 0 & 0 & \dfrac{2}{\alpha_3} & 0 & a_{35} \\ a_{41} & a_{24} &0 & \dfrac{2}{\alpha_4} & 0 \\ 0 & 0 & a_{35} &0 & \dfrac{2}{\alpha_5}  \end{array}\right){\mathbf A}
\end{equation}
\paragraph*{Class 5 : Set 11}

\begin{equation}
Q_{11}^I=\frac{1}{2}{\mathbf A}^T\left(\begin{array}{ccccc}\dfrac{2}{\alpha_1} & 0 & 2a_{13} & 0 & 0 \\ 0 & \dfrac{2}{\alpha_2} & 2a_{23} & a_{24} & 0 \\ 2a_{13} & 2a_{23} & \dfrac{2}{\alpha_3} & 0 & 0 \\ 0 & a_{24} &0 & \dfrac{2}{\alpha_4} & 0 \\ 0 & 0 & 0 & 0 & \dfrac{2}{\alpha_5}  \end{array}\right){\mathbf A}
\end{equation}
\paragraph*{Class 6 : Set 8}

\begin{equation}
Q_{8}^I=\frac{1}{2}{\mathbf A}^T\left(\begin{array}{ccccc}\dfrac{6}{\alpha_1} & 0 & a_{31} & 0 & a_{51} \\ 0 & \dfrac{6}{\alpha_2} & 0 & 5a_{24} & 0 \\ a_{31} & 0 & \dfrac{6}{\alpha_3} & 6a_{34} & 5a_{35} \\ 0 & 5a_{24} & 6a_{34} & \dfrac{6}{\alpha_4} & 0 \\ a_{51} & 0 & 5a_{35} &0 & \dfrac{6}{\alpha_5}  \end{array}\right){\mathbf A}
\end{equation}
where ${\mathbf A}$ is the column vector with entries $A_1,\ldots,A_5$ defined in \eqref{kernel1}.

With respect to the above classes, we tabulate below the couplings that have to be set to zero for each set of R-charge relations so that it admits a solution to the marginality condition:
\begin{itemize}
\item[Class 1]
In the first class of Diophantine equations one has to set the couplings of the operators depicted in Fig.~\ref{fig:cI} to zero together with the quintic operator formed by the outer pentagon of the quiver. The setting of this figure corresponds to the set 22 of \eqref{subsets1}. Then by rotating four times according to the rotational symmetries of the dihedral group on the pentagon, one gets the zero couplings corresponding to sets $25,26,27,31$ respectively.
\begin{figure}[tbh!!!]
\centering
\begin{tabular}{ccc}
\includegraphics[scale=0.20]{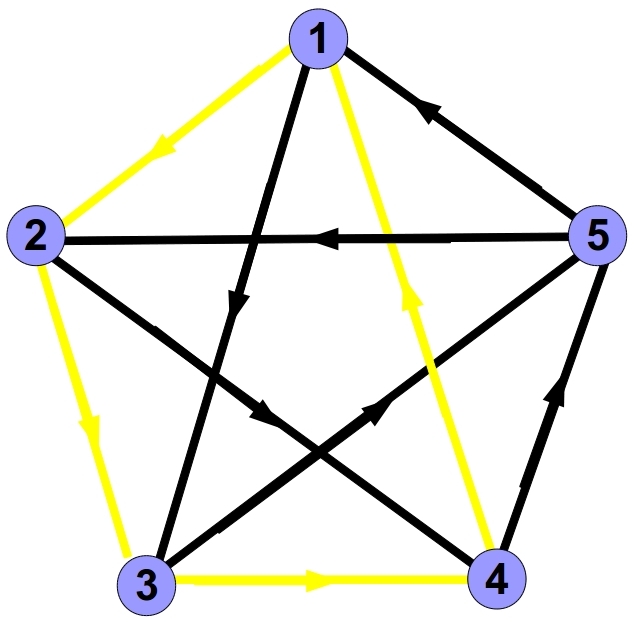} &
\includegraphics[scale=0.20]{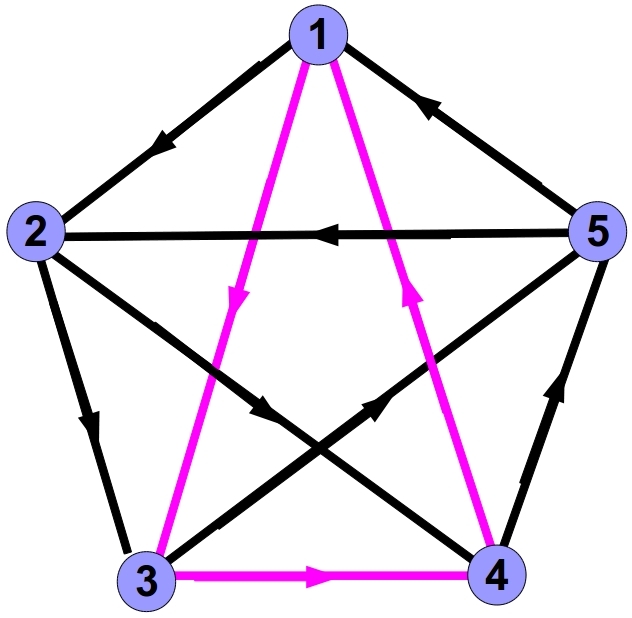} &
\includegraphics[scale=0.20]{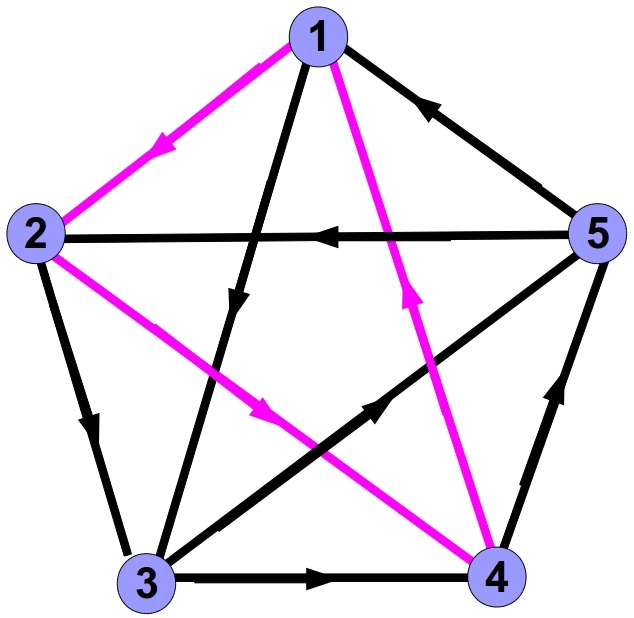}
\end{tabular}
\caption{{\sf A sketch of the operators whose couplings are set to zero. The configuration given by the superposition of the three images corresponds to the set $22$ of class $1$. Sets $25,26,27,31$ can be found by four consecutive rotations.}}
\label{fig:cI}
\end{figure}
\begin{figure}[tbh!!!]
\centering
\begin{tabular}{ccccc}
\includegraphics[scale=0.13]{pix/c1q.jpg} &
\includegraphics[scale=0.13]{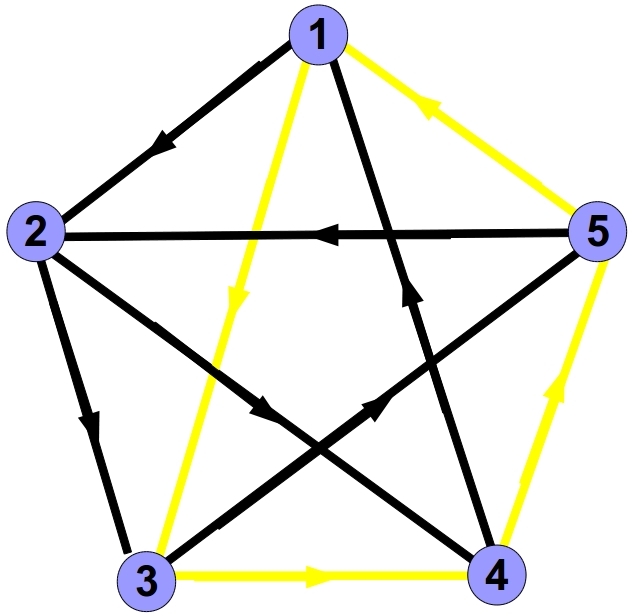} &
\includegraphics[scale=0.13]{pix/c1c1.jpg} &
\includegraphics[scale=0.13]{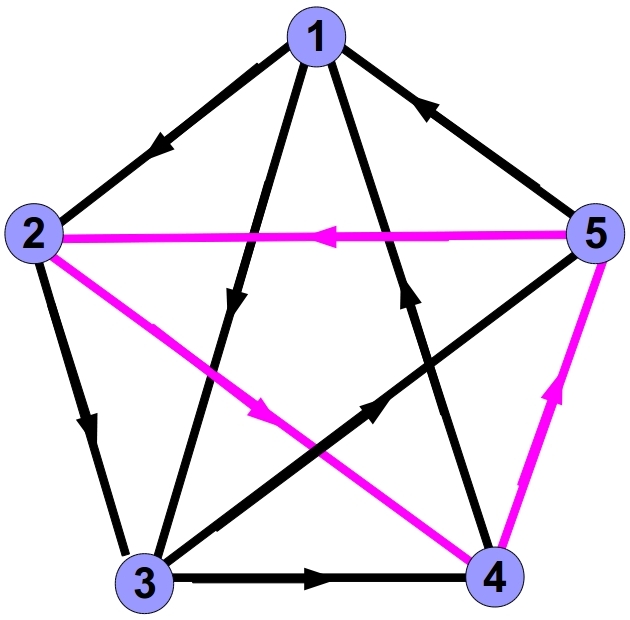} &
\includegraphics[scale=0.13]{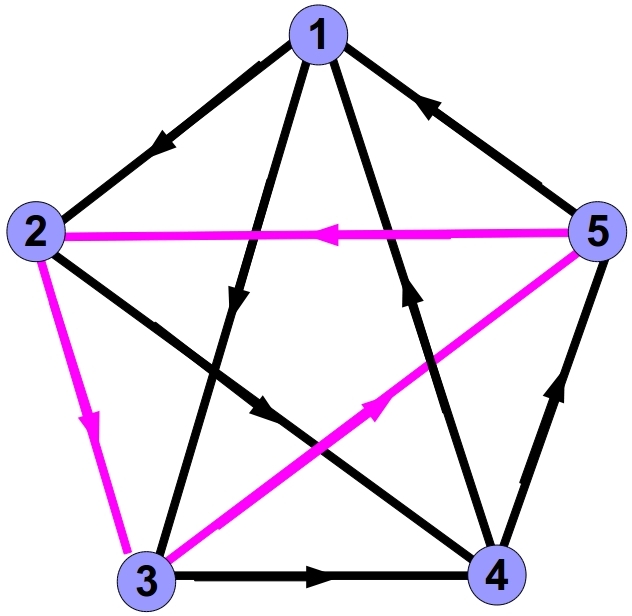} 
\end{tabular}
\caption{{\sf Set 4 of class 1. Sets 10,12,13,14 can be found by rotations.}}
\label{fig:cII}
\end{figure}
For the rest five sets of class 1 the couplings to be set to zero are depicted in Fig.~\ref{fig:cII}. This figure corresponds to set $4$ and by rotating four times one gets the couplings of sets $10,12,13,14$.
\item[Class 2]
The five equations of class 2 are described by setting to zero \comment{the quintic operator corresponding to the outer pentagon of the quiver,} the five quadratic operators and one out of five cubics each time.

\item[Class 3]
For class 3 the initial set of zero couplings, corresponding to set $23$, is depicted in Fig.~\ref{fig:cIII}. The rest can be found by rotating as previously.
\begin{figure}[tbh!!!]
\centering
\begin{tabular}{ccccc}
\includegraphics[scale=0.13]{pix/c1q.jpg} &
\includegraphics[scale=0.13]{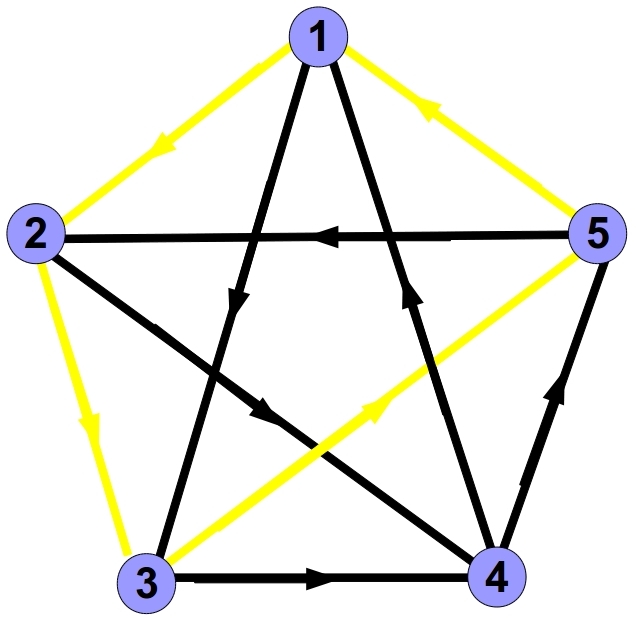} &
\includegraphics[scale=0.13]{pix/c3q3.jpg} &
\includegraphics[scale=0.13]{pix/c1c2.jpg} &
\includegraphics[scale=0.13]{pix/c3c2.jpg} 
\end{tabular}
\caption{{\sf Set 23 of class 3. Sets 24,29,30,32 can be found by rotating.}}
\label{fig:cIII}
\end{figure}

\item[Class 4]
For class 4 and set 1 we set to zero the coupling of the quintic operator formed by external lines as well as all the quadratic operators, while for set 33 we set to zero only the other quintic operator formed by the internal lines of the quiver diagram. Note that this set is the unique one with maximal cardinality. For the rest five sets of class 4 we start by the configuration of Fig.~\ref{fig:cIV} corresponding to set 2 and rotate consecutively. 
\begin{figure}[tbh!!]
\centering
\begin{tabular}{ccc}
\includegraphics[scale=0.20]{pix/c1q.jpg} &
\includegraphics[scale=0.20]{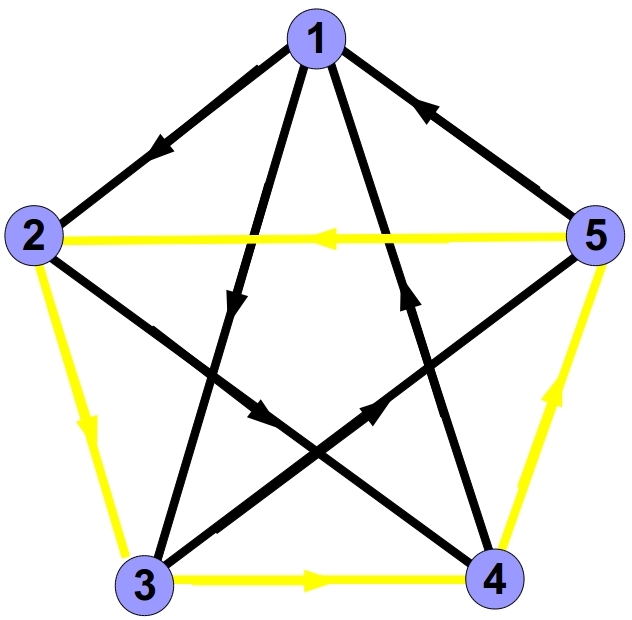} &
\includegraphics[scale=0.20]{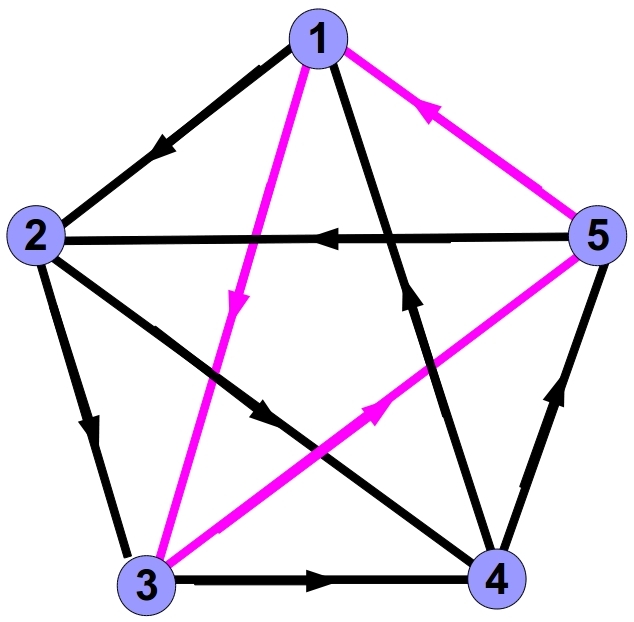} 
\end{tabular}
\caption{{\sf Set 2 of class 4. Sets 5,6,16,17 can be found by rotating this configuration.}}
\label{fig:cIV}
\end{figure}

\item[Class 5]
For class 5 the initial configuration to be rotated is depicted in Fig.~\ref{fig:cV} and corresponds to set 3.
\begin{figure}[tbh!!!]
\centering
\begin{tabular}{cccc}
\includegraphics[scale=0.15]{pix/c1q.jpg} &
\includegraphics[scale=0.15]{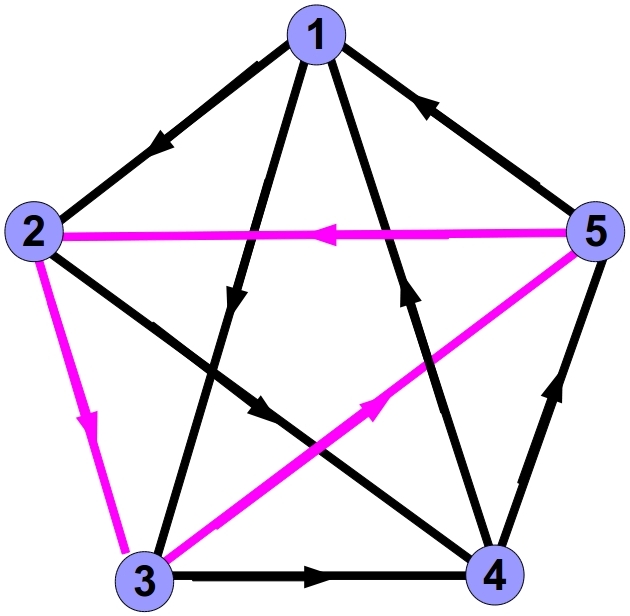} &
\includegraphics[scale=0.15]{pix/253.jpg} &
\includegraphics[scale=0.15]{pix/142.jpg}
\end{tabular}
\caption{{\sf Set 3 of class 5. Sets 7,9,11,15 can be found by rotating this configuration.}}
\label{fig:cV}
\end{figure}

\item[Class 6]
Finally, class 6 contains only one set of simultaneously marginal operators. It corresponds to setting the coupling of the ``outer" quintic operator as well as all the cubic operators to zero.
\end{itemize}
\subsection{Properties of Other Types}\label{other-types}

Here we present the structure of the superpotential through the R-charge relations for rest of the five-block quivers. 

\paragraph*{Type II, III and V}

The R-charge relations for Type II are:

\begin{eqnarray}\label{rcharges2}
r_{43} + r_{54} + r_{15} + r_{21} + r_{32} & = & 2 \nonumber\\
r_{15} + r_{53} + r_{32} + r_{21} & = & 2 \nonumber\\
r_{43} + r_{54} + r_{15} + r_{31} & = & 2 \nonumber\\
r_{43} + r_{21} + r_{32} + r_{14} & = & 2 \nonumber\\
r_{43} + r_{54} + r_{32} + r_{25} & = & 2 \\
r_{43} + r_{32} + r_{24} & = & 2 \nonumber\\
r_{15} + r_{53} + r_{31} & = & 2 \nonumber\\
r_{25} + r_{32} + r_{53} & = & 2 \nonumber\\
r_{43} + r_{31} + r_{14} & = & 2 \nonumber
\end{eqnarray}

For Type III we have:

\begin{eqnarray}\label{rcharges3}
r_{43} + r_{54} + r_{15} + r_{12} + r_{23} & = & 2 \nonumber\\
r_{15} + r_{45} + r_{43} + r_{13} & = & 2 \nonumber\\
r_{34} + r_{23} + r_{12} + r_{14} & = & 2 \nonumber\\
r_{34} + r_{45} + r_{23} + r_{25} & = & 2 \\
r_{45} + r_{43} + r_{35} & = & 2 \nonumber\\
r_{43} + r_{23} + r_{24} & = & 2 \nonumber\\
r_{43} + r_{14} + r_{13} & = & 2 \nonumber
\end{eqnarray}
and for Type V:

\begin{eqnarray}\label{rcharges5}
r_{34} + r_{45} + r_{15} + r_{12} + r_{23} & = & 2 \nonumber\\
r_{15} + r_{45} + r_{34} + r_{13} & = & 2 \nonumber\\
r_{34} + r_{45} + r_{23} + r_{25} & = & 2 \\
r_{45} + r_{34} + r_{35} & = & 2 \nonumber\\
r_{34} + r_{23} + r_{24} & = & 2 \nonumber\\
r_{15} + r_{14} + r_{45} & = & 2 \nonumber
\end{eqnarray}

These relations when more than 6 they are linearly dependent, leading to unique Diophantine equations related to the Type I special subset (33) discussed in detail in the main part of the paper.

\paragraph*{Type IV}
The R-charge relations for this type are:
\begin{eqnarray}\label{rcharges4}
r_{34} + r_{45} + r_{15} + r_{12} + r_{23} & = & 2 \label{qui41}\\
r_{34} + r_{15} + r_{13} + r_{24} + r_{25} & = & 2 \label{qui42}\\
r_{34} + r_{12} + r_{14} + r_{35} + r_{25} & = & 2 \label{qui43}\\
r_{34} + r_{35} + r_{24} + r_{25} & = & 2 \label{qua41}\\
r_{34} + r_{45} + r_{15} + r_{13} & = & 2 \label{qua42}\\
r_{34} + r_{12} + r_{23} + r_{14} & = & 2 \label{qua43}\\
r_{15} + r_{25} + r_{12} & = & 2 \label{cu41}\\
r_{34} + r_{45} + r_{35} & = & 2 \label{cu42}\\
r_{34} + r_{23} + r_{24} & = & 2 \label{cu43}\\
r_{34} + r_{13} + r_{14} & = & 2 \label{cu44}
\end{eqnarray}
Out of these 10 R-charge equations one can choose 22 sets of six linearly independent which lead to 22 Diophantine equations. 

These are:
\begin{eqnarray}
{\mathbf{\scriptscriptstyle 1}}\{\eqref{qui43},\eqref{qua41},\eqref{qua43},\eqref{cu41},\eqref{cu42},\eqref{cu44}\} & , &
{\mathbf{\scriptscriptstyle 2}}\{\eqref{qui42},\eqref{qua41},\eqref{qua42},\eqref{cu41},\eqref{cu44},\eqref{cu43}\} \nonumber\\
{\mathbf{\scriptscriptstyle 3}}\{\eqref{qui42},\eqref{qui43},\eqref{cu41},\eqref{cu42},\eqref{cu44},\eqref{cu43}\} & , &
{\mathbf{\scriptscriptstyle 4}}\{\eqref{qui42},\eqref{qui43},\eqref{qua41},\eqref{qua42},\eqref{qua43},\eqref{cu41}\} \nonumber\\
{\mathbf{\scriptscriptstyle 5}}\{\eqref{qui42},\eqref{qui43},\eqref{qua43},\eqref{cu41},\eqref{cu42},\eqref{cu44}\} & , &
{\mathbf{\scriptscriptstyle 6}}\{\eqref{qui42},\eqref{qui43},\eqref{qua42},\eqref{cu41},\eqref{cu44},\eqref{cu43}\} \nonumber\\
{\mathbf{\scriptscriptstyle 7}}\{\eqref{qui41},\eqref{qui43},\eqref{cu41},\eqref{cu42},\eqref{cu44},\eqref{cu43}\} & , &
{\mathbf{\scriptscriptstyle 8}}\{\eqref{qui41},\eqref{qui43},\eqref{qua41},\eqref{cu41},\eqref{cu42},\eqref{cu44}\} \nonumber\\
{\mathbf{\scriptscriptstyle 9}}\{\eqref{qui41},\eqref{qui43},\eqref{qua41},\eqref{qua42},\eqref{qua43},\eqref{cu41}\} & , &
{\mathbf{\scriptscriptstyle 10}}\{\eqref{qui41},\eqref{qui43},\eqref{qua42},\eqref{cu41},\eqref{cu42},\eqref{cu43}\} \nonumber\\
{\mathbf{\scriptscriptstyle 11}}\{\eqref{qui41},\eqref{qui42},\eqref{cu41},\eqref{cu42},\eqref{cu44},\eqref{cu43}\} & , &
{\mathbf{\scriptscriptstyle 12}}\{\eqref{qui41},\eqref{qui42},\eqref{qua41},\eqref{cu41},\eqref{cu44},\eqref{cu43}\} \nonumber\\
{\mathbf{\scriptscriptstyle 13}}\{\eqref{qui41},\eqref{qui42},\eqref{qua41},\eqref{qua42},\eqref{qua43},\eqref{cu41}\} & , &
{\mathbf{\scriptscriptstyle 14}}\{\eqref{qui41},\eqref{qui42},\eqref{qui43},\eqref{cu41},\eqref{cu44},\eqref{cu43}\} \nonumber\\
{\mathbf{\scriptscriptstyle 15}}\{\eqref{qui41},\eqref{qui42},\eqref{qui43},\eqref{cu41},\eqref{cu42},\eqref{cu44}\} & , &
{\mathbf{\scriptscriptstyle 16}}\{\eqref{qui41},\eqref{qui42},\eqref{qui43},\eqref{cu41},\eqref{cu42},\eqref{cu43}\} \nonumber\\
{\mathbf{\scriptscriptstyle 17}}\{\eqref{qui41},\eqref{qui42},\eqref{qui43},\eqref{qua41},\eqref{qua43},\eqref{cu41}\} & , &
{\mathbf{\scriptscriptstyle 18}}\{\eqref{qui41},\eqref{qui42},\eqref{qui43},\eqref{qua41},\eqref{qua42},\eqref{cu41}\} \nonumber\\
{\mathbf{\scriptscriptstyle 19}}\{\eqref{qui41},\eqref{qui42},\eqref{qui43},\eqref{qua41},\eqref{qua42},\eqref{qua43}\} & , &
{\mathbf{\scriptscriptstyle 20}}\{\eqref{qui41},\eqref{qui42},\eqref{qui43},\eqref{qua42},\eqref{qua43},\eqref{cu41}\} \nonumber\\
{\mathbf{\scriptscriptstyle 21}}\{\eqref{qui41},\eqref{qui42},\eqref{qua43},\eqref{cu41},\eqref{cu42},\eqref{cu43}\} & , &
{\mathbf{\scriptscriptstyle 22}}\{\eqref{qui41},\eqref{qua42},\eqref{qua43},\eqref{cu41},\eqref{cu42},\eqref{cu43}\} \nonumber
\end{eqnarray}

All subsets are related to the Type I equivalence classes.

\paragraph*{Type VI}

The R-charge relations for the last type are
\begin{eqnarray}\label{rcharges6}
r_{34} + r_{45} + r_{15} + r_{12} + r_{23} & = & 2 \label{qui16} \\
r_{12} + r_{13} + r_{35} + r_{45} + r_{24} & = & 2 \label{qui26} \\
r_{15} + r_{45} + r_{24} + r_{12} & = & 2 \label{qua16} \\
r_{25} + r_{35} + r_{12} + r_{13} & = & 2 \label{qua26} \\
r_{13} + r_{35} + r_{45} + r_{14} & = & 2 \label{qua36} \\
r_{34} + r_{45} + r_{35} & = & 2 \label{cu16}\\
r_{12} + r_{23} + r_{13} & = & 2 \label{cu26} \\
r_{15} + r_{45} + r_{14} & = & 2 \label{cu36} \\
r_{15} + r_{25} + r_{12} & = & 2 \label{cu46} 
\end{eqnarray}
Out of these equations one can pick 11 sets of 6 linearly independent,
which lead to 11 Diophantine equations. All subsets are again related to the Type I equivalence classes.
These are:
\begin{eqnarray}
{\mathbf{\scriptscriptstyle 1}}\{\eqref{qui26},\eqref{qua16},\eqref{qua26},\eqref{qua36},\eqref{cu16},\eqref{cu26}\} &,&
{\mathbf{\scriptscriptstyle 2}}\{\eqref{qui16},\eqref{qui26},\eqref{qua26},\eqref{qua36},\eqref{cu16},\eqref{cu26}\} \nonumber\\
{\mathbf{\scriptscriptstyle 3}}\{\eqref{qui16},\eqref{qui26},\eqref{qua26},\eqref{cu16},\eqref{cu26},\eqref{cu36}\} &,&
{\mathbf{\scriptscriptstyle 4}}\{\eqref{qui16},\eqref{qui26},\eqref{qua36},\eqref{cu16},\eqref{cu26},\eqref{cu46}\} \nonumber\\
{\mathbf{\scriptscriptstyle 5}}\{\eqref{qui16},\eqref{qui26},\eqref{qua16},\eqref{qua26},\eqref{qua36},\eqref{cu26}\} &,&
{\mathbf{\scriptscriptstyle 6}}\{\eqref{qui16},\eqref{qui26},\eqref{qua16},\eqref{qua26},\eqref{qua36},\eqref{cu16}\} \nonumber\\
{\mathbf{\scriptscriptstyle 7}}\{\eqref{qui16},\eqref{qui26},\eqref{cu16},\eqref{cu26},\eqref{cu36},\eqref{cu46}\} &,&
{\mathbf{\scriptscriptstyle 8}}\{\eqref{qui16},\eqref{qua16},\eqref{qua26},\eqref{qua36},\eqref{cu16},\eqref{cu26}\} \nonumber\\
{\mathbf{\scriptscriptstyle 9}}\{\eqref{qui16},\eqref{qua16},\eqref{qua26},\eqref{cu16},\eqref{cu26},\eqref{cu36}\} &,&
{\mathbf{\scriptscriptstyle 10}}\{\eqref{qui16},\eqref{qua16},\eqref{qua36},\eqref{cu16},\eqref{cu26},\eqref{cu46}\} \nonumber\\
{\mathbf{\scriptscriptstyle 11}}\{\eqref{qui16},\eqref{qua16},\eqref{cu16},\eqref{cu26},\eqref{cu36},\eqref{cu46}\} \nonumber
\end{eqnarray}

\bibliographystyle{unsrt}
\bibliography{paperBlock}

\begin{thebibliography}{10}

\bibitem{Crawley}
W.~Crawley-Boevey.
\newblock Lectures on representations of quivers.
\newblock {\em available at www.maths.leeds.ac.uk/~pmtwc/quivlecs.pdf}.

\bibitem{Derksen}
H.~Derksen and J.~Weyman.
\newblock Quiver representations.
\newblock {\em Notices Amer. Math. Soc}, 52:200--206, 2005.

\bibitem{Savage}
Alistair Savage.
\newblock Finite dimensional algebras and quivers.
\newblock {\em Encyclopedia of Mathematical Physics}, 2:313--320, 2005.

\bibitem{Brion}
Michel Brion.
\newblock Representations of quivers.
\newblock {\em Notes de l' \'ecole d' \'et\'e ``Geometric Methods in
  Representation Theory''}, 2008.

\bibitem{Assem:book}
Ibrahim Assem, Andrzej Skowronski, and Daniel Simson.
\newblock {\em {Elements of Representation Theory of Associative Algebras, Vol.
  1}}.
\newblock Cambridge University Press, 2006.

\bibitem{Douglas:1996sw}
Michael~R. Douglas and Gregory~W. Moore.
\newblock D-branes, quivers, and ale instantons.
\newblock 1996.

\bibitem{Benvenuti:2004dw}
Sergio Benvenuti and Amihay Hanany.
\newblock New results on superconformal quivers.
\newblock {\em JHEP}, 0604:032, 2006.

\bibitem{He:1999xj}
Yang-Hui He.
\newblock {Some remarks on the finitude of quiver theories}.
\newblock {\em In.J.Math.Math.Sci.}, 1999.

\bibitem{Hanany:1998sd}
Amihay Hanany and Yang-Hui He.
\newblock {NonAbelian finite gauge theories}.
\newblock {\em JHEP}, 9902:013, 1999.

\bibitem{Berenstein:2006pk}
David Berenstein and Samuel Pinansky.
\newblock {The Minimal Quiver Standard Model}.
\newblock {\em Phys.Rev.}, D75:095009, 2007.

\bibitem{Maldacena:1997re}
Juan~M. Maldacena.
\newblock The large n limit of superconformal field theories and supergravity.
\newblock {\em Adv.Theor.Math.Phys.}, 2:231--252, 1998.

\bibitem{Johnson:1996py}
Clifford~V. Johnson and Robert~C. Myers.
\newblock {Aspects of type IIB theory on ALE spaces}.
\newblock {\em Phys.Rev.}, D55:6382--6393, 1997.

\bibitem{Douglas:1997de}
Michael~R. Douglas, Brian~R. Greene, and David~R. Morrison.
\newblock {Orbifold resolution by D-branes}.
\newblock {\em Nucl.Phys.}, B506:84--106, 1997.

\bibitem{Beasley:1999uz}
Chris Beasley, Brian~R. Greene, C.I. Lazaroiu, and M.R. Plesser.
\newblock {D3-branes on partial resolutions of Abelian quotient singularities
  of Calabi-Yau threefolds}.
\newblock {\em Nucl.Phys.}, B566:599--640, 2000.

\bibitem{Feng:2000mi}
Bo~Feng, Amihay Hanany, and Yang-Hui He.
\newblock {D-brane gauge theories from toric singularities and toric duality}.
\newblock {\em Nucl.Phys.}, B595:165--200, 2001.

\bibitem{Feng:2001bn}
Bo~Feng, Amihay Hanany, Yang-Hui He, and Angel~M. Uranga.
\newblock {Toric duality as Seiberg duality and brane diamonds}.
\newblock {\em JHEP}, 0112:035, 2001.

\bibitem{Hanany:2005ve}
Amihay Hanany and Kristian~D. Kennaway.
\newblock {Dimer models and toric diagrams}.
\newblock 2005.

\bibitem{Franco:2005rj}
Sebastian Franco, Amihay Hanany, Kristian~D. Kennaway, David Vegh, and Brian
  Wecht.
\newblock {Brane dimers and quiver gauge theories}.
\newblock {\em JHEP}, 0601:096, 2006.

\bibitem{Franco:2005sm}
Sebastian Franco, Amihay Hanany, Dario Martelli, James Sparks, David Vegh,
  et~al.
\newblock Gauge theories from toric geometry and brane tilings.
\newblock {\em JHEP}, 0601:128, 2006.

\bibitem{Feng:2005gw}
Bo~Feng, Yang-Hui He, Kristian~D. Kennaway, and Cumrun Vafa.
\newblock {Dimer models from mirror symmetry and quivering amoebae}.
\newblock {\em Adv.Theor.Math.Phys.}, 12:3, 2008.

\bibitem{Hanany:2012hi}
Amihay Hanany and Rak-Kyeong Seong.
\newblock {Brane Tilings and Reflexive Polygons}.
\newblock {\em Fortsch.Phys.}, 60:695--803, 2012.

\bibitem{Cachazo:2001gh}
F.~Cachazo, S.~Katz, and C.~Vafa.
\newblock {Geometric transitions and N=1 quiver theories}.
\newblock 2001.

\bibitem{Wijnholt:2002qz}
Martijn Wijnholt.
\newblock {Large volume perspective on branes at singularities}.
\newblock {\em Adv.Theor.Math.Phys.}, 7:1117--1153, 2004.

\bibitem{Feng:2007ur}
Bo~Feng, Amihay Hanany, and Yang-Hui He.
\newblock {Counting gauge invariants: The Plethystic program}.
\newblock {\em JHEP}, 0703:090, 2007.

\bibitem{Hewlett:2009bx}
Joseph Hewlett and Yang-Hui He.
\newblock {Probing the Space of Toric Quiver Theories}.
\newblock {\em JHEP}, 1003:007, 2010.

\bibitem{Davey:2009bp}
John Davey, Amihay Hanany, and Jurgis Pasukonis.
\newblock {On the Classification of Brane Tilings}.
\newblock {\em JHEP}, 1001:078, 2010.

\bibitem{Hanany:2010cx}
Amihay Hanany, Domenico Orlando, and Susanne Reffert.
\newblock {Sublattice Counting and Orbifolds}.
\newblock {\em JHEP}, 1006:051, 2010.

\bibitem{Hanany:2010ne}
Amihay Hanany and Rak-Kyeong Seong.
\newblock {Symmetries of Abelian Orbifolds}.
\newblock {\em JHEP}, 1101:027, 2011.

\bibitem{1997alg.geom..3027K}
B.~V. Karpov and D.~Y. Nogin.
\newblock Three-block exceptional collections over del pezzo surfaces.
\newblock {\em Izv. Ross. Nauk Ser. Mat}, 62:429--463, 1998.

\bibitem{Herzog:2003dj}
Christopher~P. Herzog and Johannes Walcher.
\newblock {Dibaryons from exceptional collections}.
\newblock {\em JHEP}, 0309:060, 2003.

\bibitem{Herzog:2003zc}
Christopher~P. Herzog.
\newblock Exceptional collections and del pezzo gauge theories.
\newblock {\em JHEP}, 0404:069, 2004.

\bibitem{Herzog:2004qw}
Christopher~P. Herzog.
\newblock Seiberg duality is an exceptional mutation.
\newblock {\em JHEP}, 0408:064, 2004.

\bibitem{Feng:2002kk}
Bo~Feng, Amihay Hanany, Yang~Hui He, and Amer Iqbal.
\newblock {Quiver theories, soliton spectra and Picard-Lefschetz
  transformations}.
\newblock {\em JHEP}, 0302:056, 2003.

\bibitem{Aspinwall:2004vm}
Paul~S. Aspinwall and Ilarion~V. Melnikov.
\newblock {D-branes on vanishing del Pezzo surfaces}.
\newblock {\em JHEP}, 12:042, 2004.

\bibitem{Franco:2003ja}
Sebastian Franco, Amihay Hanany, Yang-Hui He, and Pavlos Kazakopoulos.
\newblock {Duality walls, duality trees and fractional branes}.
\newblock 2003.

\bibitem{Beasley:2001zp}
Chris~E. Beasley and M.~Ronen Plesser.
\newblock {Toric duality is Seiberg duality}.
\newblock {\em JHEP}, 0112:001, 2001.

\bibitem{Cachazo:2001sg}
F.~Cachazo, B.~Fiol, Kenneth~A. Intriligator, S.~Katz, and C.~Vafa.
\newblock {A Geometric unification of dualities}.
\newblock {\em Nucl.Phys.}, B628:3--78, 2002.

\bibitem{Fiol:2002ah}
Bartomeu Fiol.
\newblock {Duality cascades and duality walls}.
\newblock {\em JHEP}, 0207:058, 2002.

\bibitem{Berenstein:2002fi}
David Berenstein and Michael~R. Douglas.
\newblock {Seiberg duality for quiver gauge theories}.
\newblock 2002.

\bibitem{Braun:2002sb}
Volker Braun.
\newblock {On Berenstein-Douglas-Seiberg duality}.
\newblock {\em JHEP}, 0301:082, 2003.

\bibitem{Intriligator:2003jj}
Kenneth~A. Intriligator and Brian Wecht.
\newblock {The Exact superconformal R symmetry maximizes a}.
\newblock {\em Nucl.Phys.}, B667:183--200, 2003.

\bibitem{Novikov:1983uc}
V.A. Novikov, Mikhail~A. Shifman, A.I. Vainshtein, and Valentin~I. Zakharov.
\newblock {Exact Gell-Mann-Low Function of Supersymmetric Yang-Mills Theories
  from Instanton Calculus}.
\newblock {\em Nucl.Phys.}, B229:381, 1983.

\bibitem{Henningson:1998gx}
M.~Henningson and K.~Skenderis.
\newblock The holographic weyl anomaly.
\newblock {\em JHEP}, 07:023, 1998.

\bibitem{Hanany:2006nm}
Amihay Hanany, Christopher~P. Herzog, and David Vegh.
\newblock Brane tilings and exceptional collections.
\newblock {\em JHEP}, 0607:001, 2006.

\bibitem{Cecotti:2011rv}
Sergio Cecotti and Cumrun Vafa.
\newblock Classification of complete n=2 supersymmetric theories in 4
  dimensions.
\newblock 2011.

\bibitem{Alim:2011kw}
Murad Alim, Sergio Cecotti, Clay Cordova, Sam Espahbodi, Ashwin Rastogi, and
  Cumrun Vafa.
\newblock N=2 quantum field theories and their bps quivers.
\newblock 2011.

\bibitem{Cecotti:1992rm}
Sergio Cecotti and Cumrun Vafa.
\newblock {On classification of N=2 supersymmetric theories}.
\newblock {\em Commun.Math.Phys.}, 158:569--644, 1993.

\bibitem{ddpw}
Bangming Deng, Jie Du, Brian Parshall, and Jianpan Wang.
\newblock {\em Finite dimensional Algebras and Quantum Groups}, volume 150.
\newblock Mathematical Surveys and Monographs. AMS, 2008.

\bibitem{Cecotti:2012va}
Sergio Cecotti.
\newblock {Categorical Tinkertoys for N=2 Gauge Theories}.
\newblock 2012.

\bibitem{Assem:gentle}
Ibrahim Assem, Thomas Brustle, Gabrielle Charbonneau-Jodoin, and Pierre-Guy
  Plamondon.
\newblock {Gentle Algebras Arising From Surface Triangulations}.
\newblock 2009.

\end{thebibliography}
\end{document}